\documentclass[12pt,preprint]{aastex}

\shorttitle{Heavy Elements in the Solar Equation of State}
\shortauthors{Gong, D\"appen and Nayfonov}

\begin{document}

\title{Effects of Heavy Elements and Excited States in the Equation of State
of the Solar Interior}

\author{Zhigang Gong\altaffilmark{1}, Werner D\"appen\altaffilmark{1} 
and Alan Nayfonov\altaffilmark{1,2,3}}

\altaffiltext{1}{Department of Physics and Astronomy, University of Southern
California, Los Angeles, CA 90089-1342, U.S.A.}
\altaffiltext{2}{Institute of Geophysics and Planetary Physics, Lawrence
Livermore National Laboratory, Livermore, CA 94550, U.S.A.}
\altaffiltext{3}{Current Address: Teradyne Inc., 30801 Agoura Road, Agoura 
Hills, CA 91301, U.S.A.} 

\begin{abstract}
Although 98\% of the solar material consists of hydrogen and helium,
the remaining chemical elements contribute 
in a discernible way 
to the thermodynamic quantities. 
An adequate treatment
of the heavy elements and their excited states is important for 
solar models that are subject to 
the stringent requirements of helioseismology.
The contribution of various heavy
elements in a
set of thermodynamic quantities has been examined.
Characteristic features that can
trace individual heavy elements 
in the adiabatic exponent 
$\gamma_1 = (\partial\ln p /\partial\ln\varrho)_s$ (s being specific entropy),
and hence in the adiabatic sound speed were searched. It has
emerged that prominent signatures
of individual elements exist, and that
these effects are greatest in
the ionization zones, typically located near the bottom
of the convection zone.
The main result is that
part of the features found here
depend strongly both on the given species (atom
or ion) and its detailed internal partition function,
whereas other features
only depend on the presence of the species itself,
not on details such as the 
internal partition function. The
latter features are obviously well suited for a helioseismic abundance
determination,
while the former features
present a unique opportunity to use the sun as a laboratory 
to test the validity of
physical theories of partial ionization in a relatively dense and
hot plasma. This domain of plasma physics
has so far no competition from terrestrial laboratories.
Another, quite general, finding of this work is
that the inclusion of a relatively large number of 
heavy elements has a tendency to smear out individual features.
This affects both the features
that determine the abundance of elements and 
the ones that identify physical effects.
This property alleviates the task of solar modelers, because
it helps to construct a good working equation of state which
is relatively free of the uncertainties from basic physics.
By the same token, it makes more difficult
the reverse task, which is
constraining physical theories with the help of solar data.
\end{abstract}


\section{ Introduction }

The equation of state is one of the most important ingredients in solar
and stellar modeling. 
Assessing the quality of the equation of state is not easy, though.
Because 
of the relevant 
high temperatures and densities, there are no 
sufficiently accurate laboratory measurements of
thermodynamic properties that could assist the solar modeler.
However, the observational quality of helioseismology has become
so high that high-precision thermodynamic quantities must be
part of state-of-the-art solar models.
So, strictly speaking, at the
moment only theoretical studies of the equation of state
can be pursued.
But this is a too pessimistic view. Helioseismology
puts significant constraints on the thermodynamic
quantities and has already delivered powerful tools
to test the validity
and accuracy of theoretical models of the thermodynamics
of hot and dense plasmas~\citep{cd92,cd96}.

There are two important reasons why the thermodynamics can be
tested with solar data. First, the accurate solar oscillation
frequencies obtained from space and ground-based networks observation
have led to a much better understanding of the solar interior
with a very
high level of accuracy. Second, thanks to the existence of the solar
convection zone, the effect of the equation of state can be disentangled
from the other two important input-physics
ingredients of solar models, opacity and
nuclear reaction rates. This simplification is due to the fact
that in large parts of the convection zone,
the convective motion is very close to adiabatic, which leads to
a stratification that is equally very close to adiabatic. Thus, except
in the small superadiabatic zone close to the surface,
the uncertainty arising from our ignorance of the details 
of the convective motion
does not matter. As a consequence, 
inversions for the thermodynamic quantities
of the deeper layers of the convection zone promise
to be sensitive
to the small non-ideal effects employed
in different treatments of the equation of
state \citep{cd96,nd98}.

Before the 1980s, simple models of the equation of
state were used in solar modeling. Their 
physics was based on the one hand on ionization processes
modeled by the Saha equation and on the other hand on electron
degeneracy. Fairly good results had been
obtained in this way, for instance, by the equation of state of
\citet{egg73}. However, it turned out soon that the
observational progress of helioseismology had
reached such levels of accuracy that further
small non-ideal terms in the equation of state 
became observable. 
Inclusion of a
(negative) Coulomb pressure correction became imperative \citep{ulr82,
shi83,cdl88}. With this Coulomb correction being the main
non-ideal term, the subsequent
upgrade of the EFF formalism to include the Coulomb
term became quite successful; its realization is
the so-called CEFF equation of state
\citep{cd92}. Since then, other non-ideal effects, such as pressure
ionization and detailed internal partition functions of
bound states were also considered. Examples of such more advanced
equation of state are the MHD equation of state \citep{hum88,mih88,
dap88} and the OPAL equation of state \citep{rog96}, as well as the
SIREFF and its sibling, the EOS-1 equation of state 
\citep{guz97,irw01}. These efforts in basic physics have
paid off well, because a significantly better agreement between theoretical 
models and observational data has been achieved when using such 
improved equations of state \citep{cd96,bdn99}. 

Despite this progress, 
the theoretical models are not yet sufficient (see section~3.4). 
And it turns out that complicated many-body calculations are
necessary even if
the solar plasma is only slightly
non-ideal.
In recent
efforts for a better equation of
state, realistic microfield distributions \citep{nay99} and
relativistic electrons \citep{ek98,gd00,gdz00} have been 
introduced in solar models, but even these most refined
solar models have
discrepancies with respect to
the observed solar structure that are much larger
than the observational
errors themselves. 
Therefore, further investigations are still
necessary. We note in passing that while the aforementioned
non-ideal theories have all at least some theoretical footing, 
sometimes pure {\it ad-hoc}
formalisms can mimic reality even better; an example is the
recent pressure-ionization parameterization of \citet{bat00}.

The thermodynamic quantities of solar plasma do not only depend on
temperature and density, but also on the
characteristic
properties
of all atoms, ions, molecules, nuclei and electrons. 
A major complication is that the chemical composition
changes through the lifetime of the sun. Abundance changes are
a consequence of the 
several major mixing processes that have been
considered in solar models \citep{kip90}. Another change is caused by
the nuclear reactions in the solar center, which
convert four hydrogen atoms into one helium
atom, and produce, as by-products, 
several other species, such as ${}^3$He,
carbon, nitrogen and oxygen, during the p-p and
CNO chain reactions.
And finally, even in 
the solar convective zone the composition can change in time.
Although at any given moment, all elements should be
homogeneously distributed, due to the thorough stirring
of the convective motion on a very short time scale (on the order of
a month and less), over longer time scales, the chemical composition
can change nonetheless. The major effect is the so-called
gravitational
settling, which involves the depletion of heavier species 
in the stable layers
immediately below the convection zone. This depletion 
is then propagated upward into the convection zone by convective
overshoot that dips into the depleted regions.
Since the 1990s, helioseismology has been successfully
put constraints on solar models with various kinds of mixing
\citep{cox89,mv91,bp92,pro94,tbl94,ri96,tuk99},
but it is clear that a new, thermodynamics-based
determination of the local
abundance of heavy elements inside the sun would 
deliver a powerful additional constraint.

In a first part, this paper consists of a systematic study of 
the contribution of various heavy
elements in a representative 
set of thermodynamic quantities. Intuitively,
one might think that unlike in the opacity, 
in the equation of state, the influence of details in the
heavy elements is not as critical,
because the leading ideal-gas effect
of the heavy elements is
proportional to their total particle number. Their influence
should therefore 
be severely limited, since
all heavy elements together only contribute to about 
2\% in mass. However,
as shown in \citet{bat00} and in section~3.4, helioseismology has become
so
accurate that even details
of the contribution of heavy elements beyond the leading
ideal-gas term are in principle observable.
In a further part of this paper, we show
that one the one hand, there are
element-dependent, but physics-independent, features of the heavy
elements. This result has promising diagnostic possibilities for the
helioseismic heavy-element abundance determination.
On the other hand, we have also found 
features in the same thermodynamical quantities which 
do depend on the detailed physical formalism for the individual
particles. Such features will lend to a diagnosis of the physical
foundation of the equation of state.

\section{Synopsis of equation-of-state issues}

There are two basic approaches: the so-called {\it chemical} and {\it
physical} pictures. In the chemical picture, one assumes that the notion of
atoms and ions still makes sense, that is, ionization and recombination
is treated like a chemical
reaction. One of the more recent realizations of an equation of
state in the physical picture is the MHD 
equation of state. It is
based on the minimization of a model free energy.
The free energy models
the modifications of
atomic states by the surrounding plasma in a heuristic and
intuitive way, using occupation probabilities. The resulting internal
partition functions $Z_{s}^{\rm int}$ of species $s$ in MHD are

\begin{equation}\label{fo:z}
Z_s^{\rm int}=\sum_i w_{is}\,g_{is}\,\exp [-(E_{is}-E_{1s})/(k_{\rm{B}}T)] \ ,
\end{equation}

Here, $is$ label states $i$ of species $s$. $E_{is}$ are their energies, and
the coefficients $w_{is}$ are the occupation probabilities that take into
account charged and neutral surrounding particles. In physical terms, $w_{is}$
gives the fraction of all particles of species $s$ that can exist in state $i$
with an electron bound to the atom or ion, and $1 - w_{is}$ gives the fraction
of those that are so heavily perturbed by nearby neighbors that their states
are effectively destroyed. Perturbations by neutral particles are based on an
excluded-volume treatment and perturbations by charges are calculated from a
fit to a quantum-mechanical Stark-ionization theory. Hummer \& Mihalas's
(1988) choice had been

\begin{equation}\label{fo:w}
\ln w_{is}=-\Big(\frac{4\pi }{3V}\Big)\left\{{\sum_{\nu } }N_\nu
(r_{is}+r_{1\nu })^3+16 \left[\frac{(Z_s+1)e^2}{\chi _{is}k_{is}^{1/2}}\right]^3
\sum_{\alpha \neq e}N_\alpha Z_\alpha ^{3/2}\right\} \ ,
\end{equation}

Here, the index $\nu$ runs over neutral particles, the index $\alpha$ runs
over charged ions (except electrons), $r_{is}$ is the radius assigned to a
particle in state~$i$ of species~$s$, $\chi_{is}$ is the (positive) binding
energy of such a particle, $k_{is}$ is a quantum-mechanical correction, and
$Z_s$ is the net charge of a particle of species $s$. Note that $ \ln w_{is}
\propto \ -n^6 $ for large principal quantum numbers $n$ (of state $i$), and
hence provides a density-dependent cut-off for $Z_{s}^{\rm int} $.

The physical picture provides a systematic method to include nonideal effects.
An example is the OPAL equation of state, which
starts out from the grand canonical ensemble of a system of the basic
constituents (electrons and nuclei), interacting through the Coulomb
potential. Configurations corresponding to bound combinations of electrons and
nuclei, such as ions, atoms, and molecules, arise in this ensemble naturally
as terms in cluster expansions. Effects of the plasma environment on the
internal states, such as pressure ionization, are obtained directly from
statistical-mechanical analysis, rather than by assertion as in the chemical
picture.

Although the stellar plasma we deal with is assumed to be electrically
neutral in a large volume, it is a mixture of charged ions and electrons
inside. The Coulomb force between these charged particles is long range. By
doing the first order approximation, the so-called \citet{dh23}
potential results. It is the approximation of the static-screen
Coulomb potential (SSCP), which 
describes the interaction between
charged particles as

\begin{equation}\label{eq:scr}
\psi (r)=Ze\,\frac{\rm{e}^{-r/\lambda_{\rm{D}}}}{r} \ ,
\end{equation}

where $Z$ is the net charge of the particle, $e$ is electron charge, $r$ is
the distance to the center of the target particle, and $\lambda _{\rm{D}}$
is the Debye length. The corresponding free-energy, which is
sometimes called the Debye-H\"{u}ckel free energy, can then be
obtained.
It has become widely used 
\citep{gra69, mih88}. Furthermore, 
in order to eliminate the short-range divergence in 
the Debye-H\"{u}ckel potential, a cutoff function

\begin{equation}
\tau (x)=3x^{-3}\,[\ln (1+x)-x+\frac 12x^2]=3x^{-3}\,\int_0^x\frac{y^2}{1+y}
\rm{d} y \ ,
\label{Fo:tau}
\end{equation}

was introduced, where $x=r_{\min }/\lambda_{\rm{D}}$, with $r_{\min }$
being the distance of closest approach (the minimum distance that
the center of two particles can reach). This factor $\tau$ 
is essential to get rid of the possibility of a negative
total pressure when density of the plasma is becoming high. However, 
it has become known for some time \citep{bat95, dap96}
that this $\tau $ factor (Eq. \ref{Fo:tau}) can 
produce some unforeseen and unjustified effects on some
thermodynamic quantities in solar equation of state, which we will discuss
in more detail in Sect. 4. In order to clearly disentangle the
effects of the heavy elements and those of the $\tau$ factor,
in this paper we are using the MHD and CEFF equations of state
with a Debye-H\"{u}ckel theory {\it but without a $\tau$ correction}
unless explicitly stated.

In helioseismic inversions for equation-of-state effects,
the resulting natural second-order thermodynamic quantity is
the adiabatic gradient~\citep{bach97}, which is itself 
closely related to adiabatic sound speed, which can also be
the result of inversions. Such inversions are called ``primary''.
However, in the diagnosis of physical effects, other second-order 
thermodynamic quantities can in principle be more revealing
than the adiabatic gradient~\citep{nd98}. Unfortunately,
the helioseismic inversion for these other thermodynamic quantities
is less direct, because
additional physical
assumptions must be made. Such inversions are therefore 
called ``secondary''. 
In view of future primary and secondary inversions,
we study here not only the adiabatic gradient, but more
systematically the complete set of the following three second-order
thermodynamic quantities

\begin{equation}
\chi _\rho =(\partial \ln p/\partial \ln \rho )_T \ ,
\end{equation}

\begin{equation}
\chi _T=(\partial \ln p/\partial \ln T)_\rho  \ ,
\end{equation}

\begin{equation}
\gamma _1=(\partial \ln p/\partial \ln \rho )_s  \ ,
\end{equation}

where $p$ is pressure, $\rho $ is density, $T$ is temperature, and $s$ is
specific entropy. All 
other second-order quantities can be derived as functions
of these three quantities.

\section{Effects of heavy elements}
\subsection{Difference between selected equations of state}

It is well known that the heavy-element abundance of stars
is crucial for opacity, where a certain single heavy element often leads
to the dominant contribution. However, one would
expects a much less dramatic
effect in the equation of state. Quantitatively,
in the sun, the abundance of heavy-elements 
is small (about 2 percent in
mass) and their influence in the equation of state is roughly
proportional to their number abundance, which is about an order of
magnitude smaller. Nevertheless, helioseismology
has already revealed heavy-element effects~\citep{dap93, bat00}.

In this paper, we study 
the effect of the heavy 
elements with the help of 
the MHD equation of state. We choose the MHD equation of state, 
because it contains relatively detailed physics,
has been widely used in solar modeling (thus giving a benchmark
for our analysis), and most importantly, because 
among all the non-trivial equations of state
it is the only one with which a systematic study
of the influence of individual elements and their detailed physical
treatment can be carried out. The other non-trivial equations
of state are only available in pre-computed tabular form,
with fixed heavy-element abundance.

The physical conditions for which 
the different equations of state have been calculated
are from a
solar model [model S of \citet{cd96}], restricted to the
convection zone, that is, the
range relevant for a helioseismological equation-of-state diagnosis.
For our conceptual and qualitative study we do not need
the most up-to-date values for the chemical composition. Thus, for
simplicity, we have chosen a typical
chemical composition of mass fractions $X=0.70$, $Y=0.28$
and $Z=0.02$. 
Since we are also 
comparing our result with the OPAL equation of state [which is
so far best for solar models~\citep{cd96}], for the sake
of consistency, we have
chosen a distribution of heavy elements in exactly the
same way as in the OPAL tables 
\citep{rog96}. For convenience, we here list this choice
in Table~\ref{Tb1}. 

\placetable{Tb1}

\begin{figure}
\figcaption{Absolute values of thermodynamic quantities 
for different equations of state with a regular 6-element mixture
\citep{rog96} for conditions of the solar
convection zone. 
Dashed line: MHD; Thick Solid line: OPAL; 
Dashed-dotted line: CEFF; Dotted line: SIREFF; Thin Solid line: MHD$_{\rm GS}$.
\label{Fig1}}
\plotone{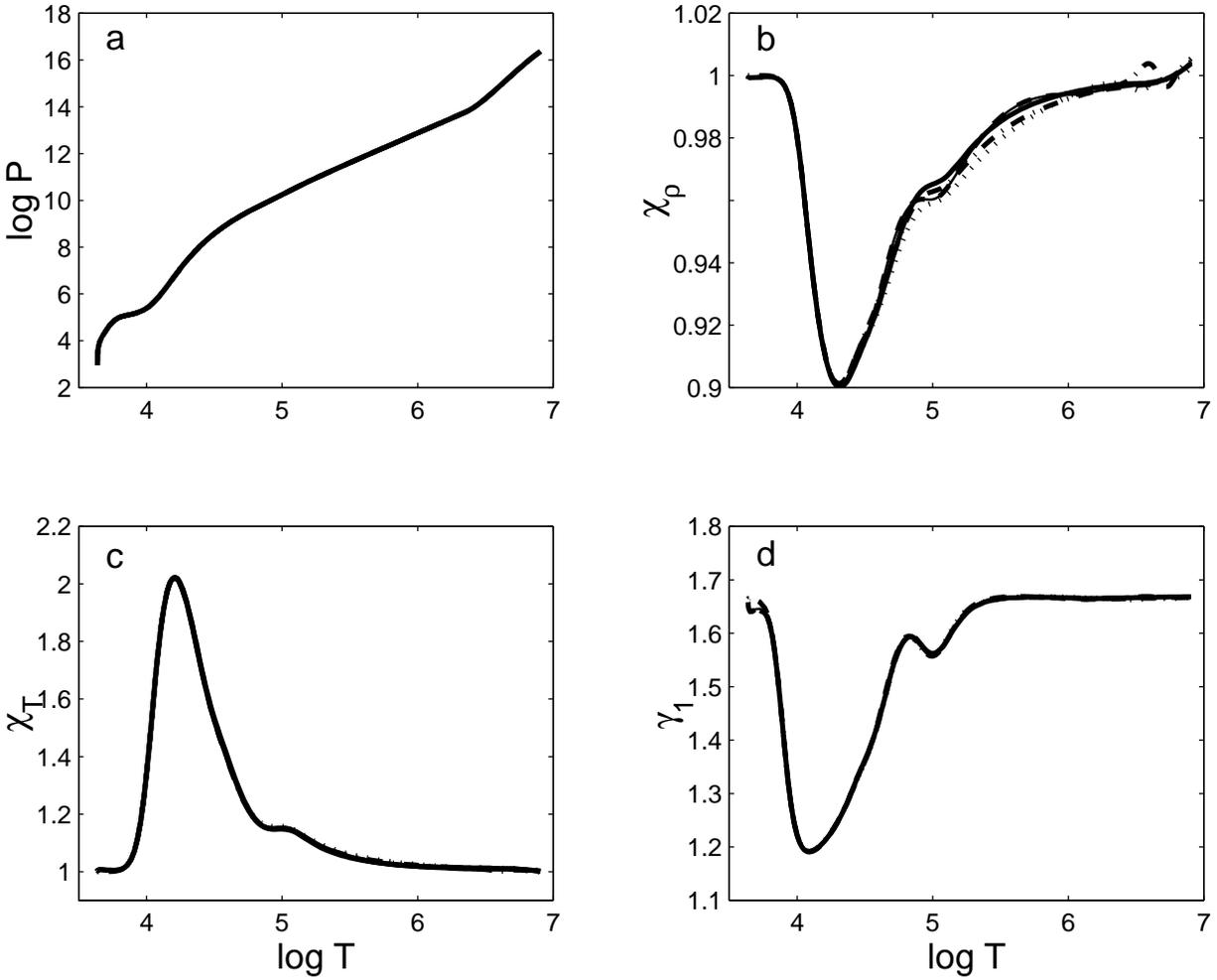}
\end{figure}

\placefigure{Fig1}

\begin{figure}
\figcaption{Relative difference in
thermodynamic quantities of Fig.~1 in the sense
[(X - X$_{\rm MHD}$)/ X$_{\rm MHD}$]. 
Dashed line: MHD; Solid line: OPAL; Dashed-dotted line: CEFF; 
Dotted line: SIREFF.
\label{Fig2}}
\plotone{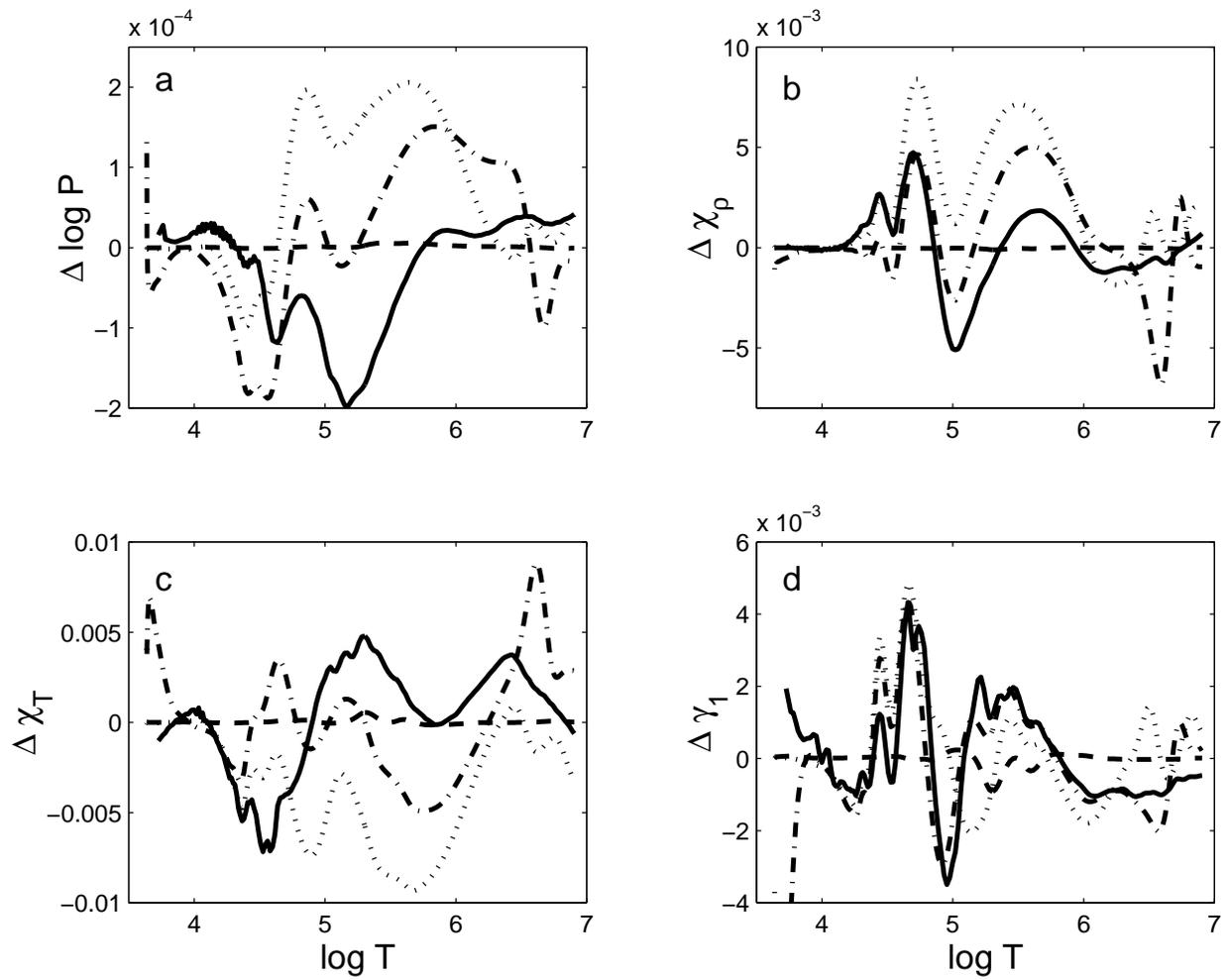}
\end{figure}

\placefigure{Fig2}

In Figs.~\ref{Fig1} and~\ref{Fig2} we show pressure, 
$\chi _\rho $, $\chi _T$
and $\gamma _1$ in a solar model with the 6-element mixture 
of~Table~\ref{Tb1}
for several popularly used equations of state. Both absolute values,
and for finer details, relative differences with
respect to the MHD equation of state are displayed. 
The equations of state are:\\
MHD - standard MHD with the usual occupation probabilities~\citep{hum88}.\\
MHD$_{\rm{GS}}$ - standard MHD but with internal partition functions of heavy
element truncated to the ground state term (however, 
the internal partition function
of H and He are not truncated, which is different from \citet{nd98} [ND98
hereafter] and the H-He mixture models of section~4).\\
OPAL - interpolated from the OPAL tables \citep{rog96} [except for the 
H-He-C mixture models, 
which were directly calculated by one of us (AN)].\\
CEFF - \citet{cd92}\\
SIREFF - \citet{guz97}\\
As found by ND98, for a pure hydrogen and a hydrogen-helium mixture under
solar conditions, among the possible thermodynamic quantities (that
derive
from the second derivatives of the free energy),
it is the quantity $\chi_\rho $ [Fig. \ref{Fig1} (b)] which reveals most
physical effects, and this already in the absolute values. 
Of course, other thermodynamic
quantities contain signatures of the same physical effects
as well, but often they would show up only in the {\it de-facto}
amplification of relative
differences.
The sensitivity of $\chi_\rho $ is mainly due to the fact 
that in ionization zones it varies considerably less
than the other thermodynamic quantities. 
Finer effects are therefore overlaid on a smaller global variation and
show therefore up already in the absolute plots. Specifically, 
the main result of ND98
was a ``wiggle'' in $\chi_\rho $, resulting from the density dependent
occupation probabilities for the excited states of hydrogen in MHD. 

Our study deals with the
combined effect of several different heavy elements and, importantly,
several different physical mechanisms that describe 
the interaction between the
ground and excited states of their atoms and ions with
the surrounding plasma.
In order to disentangle effects from individual elements,
we have studied the heavy-element effects separately for each element.
Before we interpret the heavy-element features in Fig. \ref{Fig1},
first we should like to mention an effect
not related to heavy elements, which nonetheless demonstrates
the power of our analysis.
Fig. \ref{Fig1} (d) shows, at very low 
temperature ($%
\log T\leq 3.5$), a feature. For instance, in $\gamma _1$ 
a dip appears in most equations of state
except in CEFF. It is obviously 
the signature of H$_2$ molecular dissociation \citep{ld88}, a process not
included in the CEFF equation of state.

\begin{figure}
\figcaption{Difference in thermodynamic quantities 
between the 
6-element mixture and the hydrogen-helium mixture for various
equations of state. Difference are in the sense [X(Model)$_{\rm 6-element}$
- X(Model)$_{\rm H-He}$]. Line styles the same as in Fig. \ref{Fig1}.
\label{Fig3}}
\plotone{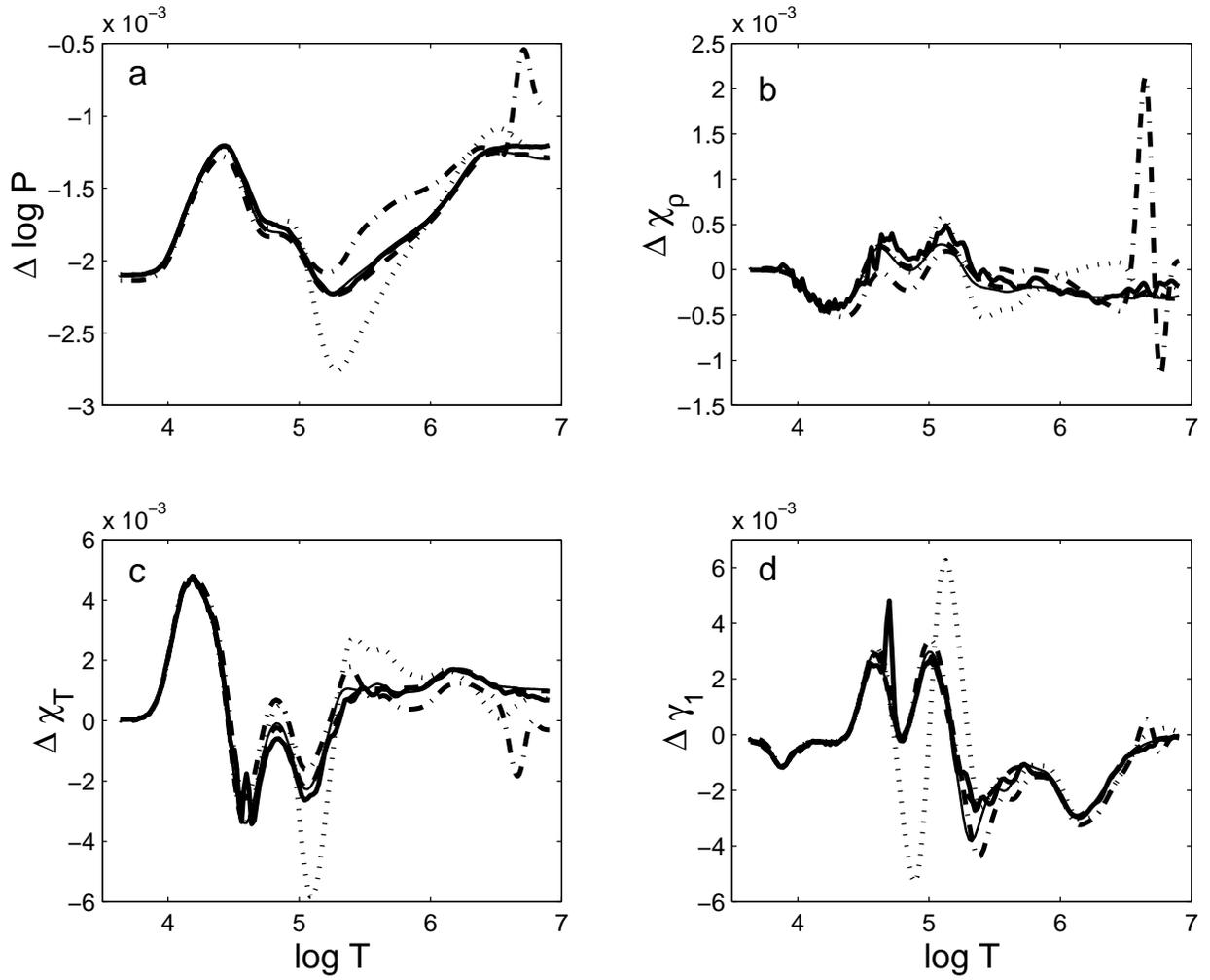}
\end{figure}

\placefigure{Fig3}

Next, we have
compared the contribution of heavy elements in the various equations of
state (Fig. \ref{Fig3}). For each particular equation of 
state, we have computed 
the relative difference between the regular 6-element mixture and a
hydrogen-helium mixture (mass fractions being $X=0.70$, $Y=0.30$).
This difference is much smaller than the one given in Fig. \ref{Fig2},
suggesting that the biggest
difference among different models of equation of state is related to the
treatment of hydrogen and helium. However, as shown below, 
this does not mean that
for second-order thermodynamic quantities
in solar models of helioseismic precision 
the contribution of heavy elements would be negligible
(see section~3.4).

\placefigure{Fig4}

\subsection{Signature of individual heavy elements}

\begin{figure}
\figcaption{Difference in thermodynamic quantities 
between the hydrogen-helium-and-one-heavy-element mixture and the 
hydrogen-helium mixture 
for the MHD equation of state. Differences are in the sense
[X$_{\rm H-He-Z}$ - X$_{\rm H-He}$]. Solid line: H-He-C; 
Dashed-dotted line: H-He-N; Dotted line: H-He-O; Dashed line: 
H-He-Ne.
\label{Fig4}}
\plotone{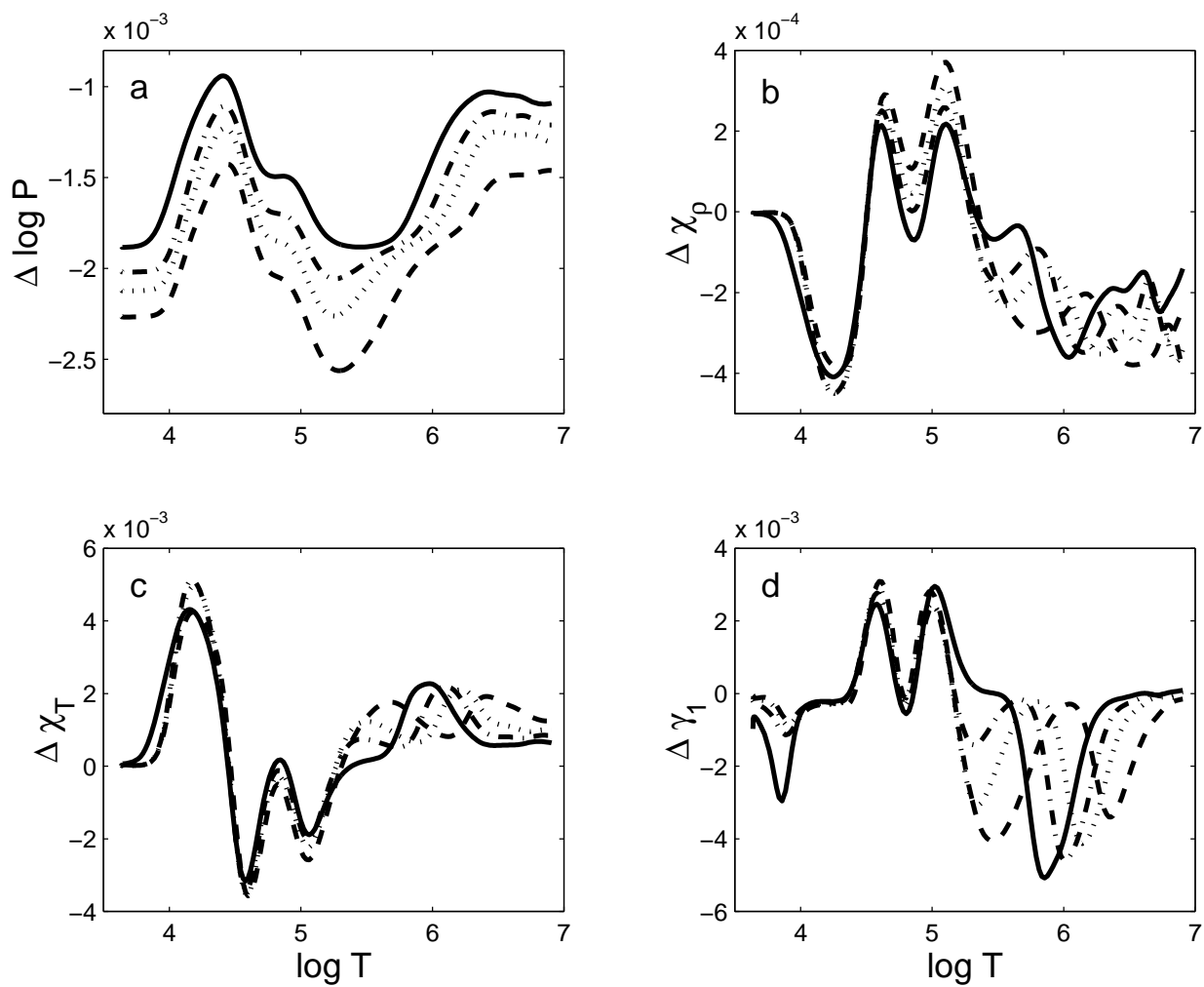}
\end{figure}

\begin{figure}
\figcaption{Difference in the 
total number of particles between the
hydrogen-helium-and-one-heavy-element mixture and the hydrogen-helium 
mixture for the MHD equation of state. Differences are in the sense
[N$_{\rm H-He-Z}$ - N$_{\rm H-He}$]. Thin Solid line: 6-element mixture;
otherwise the same as in Fig. (\ref{Fig4}).
\label{Fig5}}
\plotone{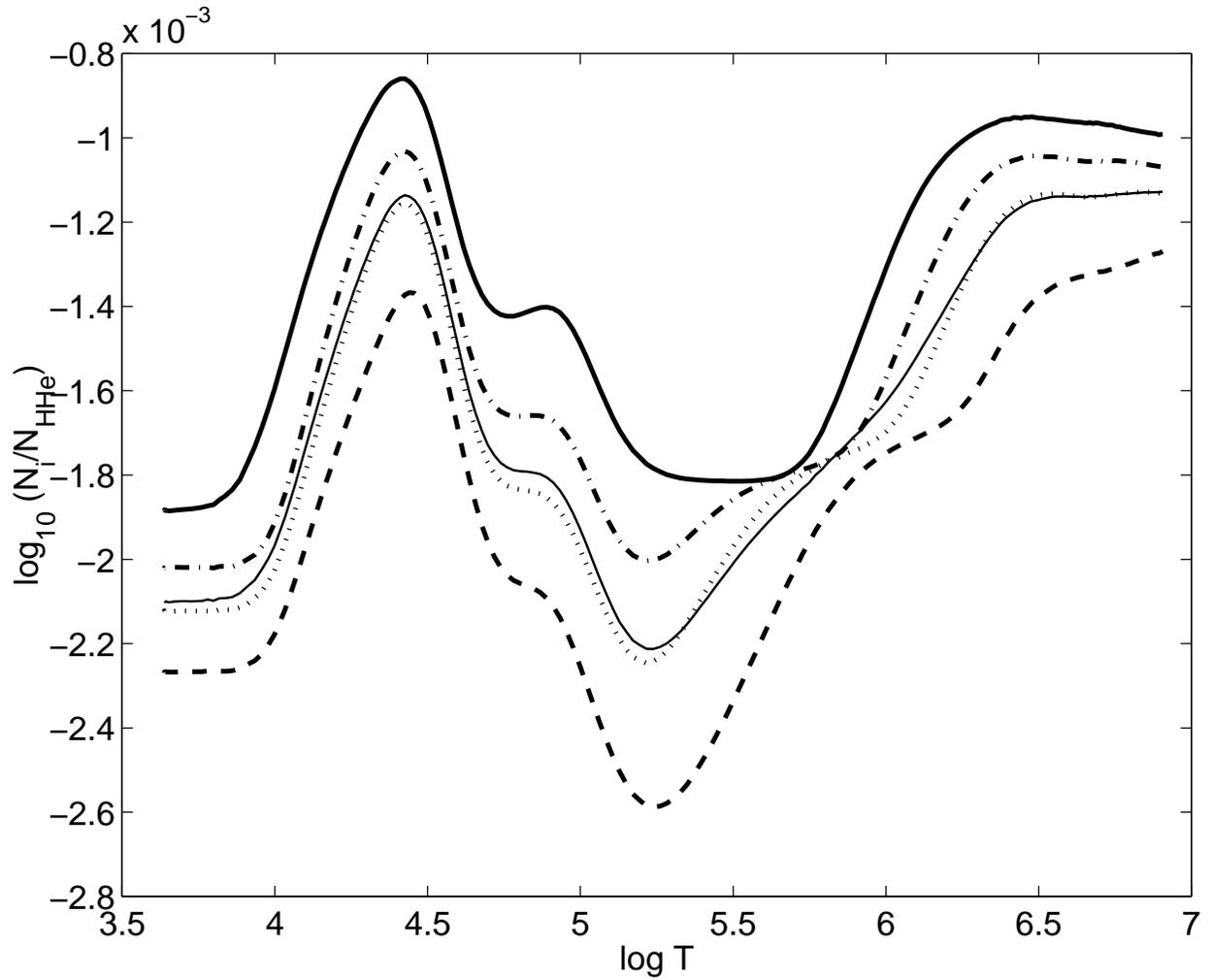}
\end{figure}

In order to reveal the contribution of each individual heavy element, we
have calculated 
solar models with particular mixtures. In each case, 
hydrogen and helium
abundances were fixed with mass fractions $X=0.70$ and $Y=0.28$;
the remaining 2\% heavy-element contribution was topped off by 
only one element, 
carbon, nitrogen, oxygen and neon,
respectively.  We have then compared 
these models with the hydrogen-helium mixture model in
Fig. \ref{Fig4} (this mixture is obtained 
by filling the two percent
reserved for heavy elements with additional helium).
The expectation is that the biggest deviations between these special
models, and the one with the complete heavy-element mixture, will result 
(i) from
the change in the total number of particles per unit volume
and (ii) from the different ionization potentials of
the respective elements.
Because the solar plasma is only slightly non-ideal,
the leading pressure term is given by the ideal-gas equation

\begin{equation}
pV=Nk_BT \ ,
\end{equation}

with $p$ standing for pressure, $N$ the
number of particles, and $k_B$ the Boltzmann constant. Because of their
higher mass, the number of heavy-element atoms is obviously smaller
than that of helium atoms representing the same mass fraction.
However, against this reduction in total number there is an offset
due to the ionization of their larger number of electrons which becomes
stronger at higher temperatures.
The net change of the total number of particles is therefore a 
combination of
these two effects. In 
Fig. \ref{Fig5} we show the difference between the total number of
particles in these models, and we see that indeed it explains the 
difference in total pressure well.

\placefigure{Fig5}

Since the second-order thermodynamic quantities (Eq.~5-7)
[see Fig. \ref{Fig4} (b-d)]
are independent of the total number of particles, they can
reveal signatures of the internal
structure of each element more directly. 
Therefore, at low temperatures ($\log T<5.2$), the
difference with respect to a hydrogen-helium mixture plasma is similar
for all the heavy elements,
because the difference in the second-order quantities 
mainly comes from the fact that the 2\% helium in the H-He mixture 
are already partly ionized, while the replacing heavier elements
are still mainly neutral.
However, as temperature rises, a signature of the ionization
of the individual heavy element appears. 
This selective modulation of second-order thermodynamic
quantities, as well as another property discussed 
further below, allow us in principle to identify single heavy elements
in the solar convection zone, in analogy to optical spectroscopy.

Our next comparison has the purpose to
disentangle even further the contribution of the mere presence
of each heavy element from 
the more subtle influence of different physical formalisms.
In Fig.
\ref{Fig6}
the H-He-C models (with mass fractions 0.70:0.28:0.02)
are compared with the H-He models (0.70:0.30) for our set of
different
equations of state. 
This procedure eliminates most of the difference due to
the treatment of hydrogen and helium in each individual 
equation of state and therefore isolates 
the behavior of carbon in each of our 
equations of state. We note that here, roughness appears in
the OPAL
equation of state, 
because the sensitivity of our analysis 
has almost reached the reasonable
limit of accuracy that can be obtained by interpolation
in the OPAL tables.

\placefigure{Fig6}
\placefigure{Fig7}

From Fig. \ref{Fig6} (a), we can see that 
among the thermodynamic quantities,
$\Delta \chi _T$ reveals the biggest differences between the equations
of state in the temperature range of
$4.5<\log T<5.5$ (named region ``A'' in the figure). 
Common features can be identified 
when $\log T>5.5$ and $\log T<4.5$.  
Such features are also visible in the 
$\Delta \gamma _1$ graph between H-He-C and H-He shown in
Fig. \ref{Fig6} (b), and between H-He-C and the 
regular 6-element mixture shown in Fig. \ref{Fig7}. 

In the region ``A'' of Fig. \ref{Fig7} 
the MHD and OPAL results together differ
significantly from the other three formalisms; 
they appear
closer to the reference 
6-element mixture model than to the other models. One obvious reason is that 
neither MHD$_{\rm{GS}}$ nor CEFF nor SIREFF include excited states of 
heavy elements (in this figure carbon), while both MHD and OPAL do,
although OPAL does it quite differently \citep{ro86}.
Our figure would then suggest 
that the difference between MHD and OPAL on the one hand,
and the other three formalisms on the other hand, appears most likely 
due to
the contribution 
of the excited states of carbon. In addition, it also appears that 
that it matters less that excited
states are treated differently than that they are not neglected.
It will be a challenge for
helioseismology to distinguish between such small differences.

The bump in $\Delta \gamma _1$ in the region ``B'' 
of either
Fig. \ref{Fig6} (b) or Fig. \ref{Fig7} is totally
independent of the equation of state used. 
By comparing with Fig. \ref{Fig4} 
(d), it follows that the
feature is likely due to the ionization of carbon at that 
temperature in general, quite 
independent of details in the equation of state. 
The strength and robustness of this profile
and its relative independence on the
equation of state qualify it ideally for
a helioseismic heavy-element abundance
determination. This is true because the
profile does not only appear for carbon, but also for all
other heavy elements which exhibit a similar feature independent of
the equation of state.
As an example, the analogous phenomenon for nitrogen 
at its higher ionization temperature is
revealed by the H-He-N model in Fig. 
\ref{Fig8}.
It is clear from Fig. \ref{Fig4} (d) that each heavy element has its own 
profile, and in analogy to the helioseismic helium-abundance
determination \citep{von94, bat00}, these profiles promise to be
used in
future inversions of solar oscillation frequencies to determine the
abundance of the heavy-elements.

\placefigure{Fig8}

\subsection{Effects of the number of heavy elements in the equation of state}

Another important issue in the study of heavy-element effects in the solar
equation of state is how many elements should be considered, that
is,
what is the error if not enough elements are included. 
We address this
question by adding species one by one until we reach
the 15 element mix considered in the original 
MHD equation of state \citep{mih90} to see if there is
subset that is adequate for helioseismic accuracy.
To be more specific, we have again
set the total abundance of all heavy elements to be fixed at
$Z=0.02$, but this time we follow the
\citet{gre96} abundance, which slightly differs from the simplified
values of
Table~\ref{Tb1}. 
We label the mixtures by the number of species 
including H and He 
(thus a 3-element mixture means H-He-C;
a 4-element H-He-C-N, {\it etc.}). 
To reconcile the absolute mass fractions of the
\citet{gre96} mixture with our specification of
a fixed mass fraction $Z=0.02$, some form of
topping off is necessary.
In the case
of the first 6 elements, we have chosen to readjust
the last element 
to top off to $Z=0.02$, with the other
heavy elements being set to the \citet{gre96}
abundance. The same convention holds for the 7th element, which
is iron. However, from this case onward to the full 15-element mixture, we
have always used iron to top off to $Z=0.02$,
with the other heavy elements always having their \citet{gre96}
abundance.

\placefigure{Fig9}

The results of this systematic procedure to add heavy elements one by
one are shown in Fig. \ref{Fig9}. 
We distinguish two cases.
First, regarding pressure, the larger
the number of
heavy elements considered is, the closer pressure approaches that of
the full mixture. This is easy to understand from
the role of the effect of individual
heavy elements discussed in Sec. 3.1. Second, and more interesting,
regarding the second-order thermodynamic quantities, the larger
the number of
heavy elements considered is, the smoother the curves overall become. 
The reason is that when more species are included the less weight
each individual species obtains and thus its contribution with its
specific features becomes reduced.
In addition, different species have their own profile which sometimes
leads
to partial cancellations.
Overall, then, the effect of a mixture with a larger number
of heavy elements is manifested by the relatively flat profiles of
Fig. \ref{Fig9} (b-d).
Third, regarding the minimum number of heavy elements necessary
for accurate helioseismic studies, we conclude
that the inclusion of new species of heavy elements becomes 
less and less critical if at least ten of the most 
abundant species are included. Inclusion of more elements will
lead to such small differences in the equation of state that they appear
to be undetectable by current helioseismological studies 
(see the following
section).
Fourth, however, the currently popular 6-element mixture used in 
OPAL is still inadequate in as far as it leads to a deviation with
respect to the full mixture, attaining up to $3 \times 10^{-4}$
in $\gamma_1$ at the base on solar convection zone.
And since the OPAL data are so far only available in tabular form,
the inevitable interpolation error, which is typically of the same order,
only aggravates this situation.

\subsection{Current resolution power of helioseismology}

The observational resolution of helioseismic inversions
is demonstrated in Fig. \ref{Fig10}. It
shows the result of an inversion~\citep{bdn99}
for the intrinsic $\gamma_1$ difference between
the sun and a solar model \citep{bach97}. 
The intrinsic difference
is the part of the $\gamma_1$ difference due
to the difference in the equation of state itself [and not the
additional part implied by the change in solar model due to that of the
equation of state, which induces a further $\gamma_1$ difference
\citep{bach97}].
The main result of
Fig. \ref{Fig10} is that 
present-day helioseismology has reached an accuracy
of almost $10^{-4}$ for $\gamma_1$. The
effects discussed in this paper are therefore within
reach of observational diagnosis. 

More specifically, studies such as the one shown
in Fig. \ref{Fig10} reveal
the influence of the equation of state by an analysis 
of the difference
between the solar values obtained from inversions and 
the ones computed in reference models (standard solar models).
In the case of Fig. \ref{Fig10}, two sets of reference
models are compared to the solar data, one set based on 
the MHD equation of state, the other set on OPAL.
Because of the differential nature of these inversions, they
become more reliable when the reference model is
close to the real solar
structure. 
In the last years, the solar models were significantly improved,
thanks to the constraints of helioseismology. In particular,
diffusion has now become part of the standard solar
model \citep{cd96} and it was included in the calibrated reference
models of~\citet{bdn99} (models M1--M8 of 
Table~\ref{Tb2}). Their composition profiles (in particular
the surface helium abundance $Y_{\rm s}$) were those
obtained from helioseismological inversions
by~\citet{anchi98}. All models had $Z/X=0.0245$
\citep{gn93}. In the inversions,~\citet{bdn99}
tested the robustness of the inferred results against
uncertainties in the solar-model inputs using a number of different solar
models (see their paper for details of the tests made).  All of the
models M1--M8 had either the MHD or the OPAL equation of state, and
they were all using OPAL opacities \citep{ir96}, supplemented
by the low temperature opacities of~\citet{ku91}. 
Since one of the most uncertain aspects
of solar modeling is always the formulation of the convective flux,
two different formalisms were used -- standard mixing length theory
(MLT) and the~\citet{cm91} formalism (CM). 
The two formalisms give fairly different stratifications in the
outer regions of the sun.
Fig.~\ref{Fig10} clearly shows that helioseismology has
the potential to address the small effects from heavy
elements such as discussed in Fig.~\ref{Fig7} and Fig.~\ref{Fig8}.

\placefigure{Fig10}

\placefigure{Fig11}

While Fig.~\ref{Fig10} is based on numerical inversions, asymptotic
inversion techniques can shed light from a different angle.
In particular, they are well
suited for heavy-element abundance determinations, as demonstrated by
Fig.~\ref{Fig11} from~\citet{bat00}. Fig.~\ref{Fig11}
shows the result of an {\it asymptotic} inversion 
introduced by~\citet{go84} for the helioseismic
helium abundance determination. The inversion is for the so-called
quantity $W$, which is a function of
solar structure

\begin{equation}
W={{\rm d}c^2\over {\rm d}r}{1 \over g(r)} \  .
\end{equation}

Here,
$g(r)$ is the local
gravitational acceleration at position $r$ in the sun.
The quantity $W$ is useful because of its equality with
a purely
thermodynamic quantity if the stratification is assumed to be perfectly
adiabatic (otherwise, the equality is violated by the amount of
the non-adiabaticity). For adiabatic stratification, the
following relation holds \citep{go84}

\begin{equation}
W={1 - {\gamma_1} - {\gamma}_{1,\rho} \over 1 - {\gamma}_{1,c^2}} \ ,
\end{equation}

with the derivatives

\begin{equation}
\gamma_{1,\rho} = \left({\partial \ln \gamma_1\over \partial \ln \rho}
\right)_{c^2}, \gamma_{1,c^2}= \left({\partial \ln \gamma_1\over
\partial \ln c^2} \right)_{\rho} \ .
\end{equation}

While the reader is referred to~\citep{bat00} for more detail, here we 
merely mention that
similarly to Fig.~\ref{Fig10}, Fig.~\ref{Fig11} shows
$W$ in the sun and in two artificial solar models, but 
in contrast to Fig. \ref{Fig10}, the inversions of Fig.~\ref{Fig11}
are
absolute and not relative to a reference model. The comparison
models of Fig.~\ref{Fig11} are based on
the H-He-C composition of section~3.2 (where 
pure carbon is representing the total heavy-element
abundance), and the analogous H-He-O composition, respectively.
It is no surprise that since 
the quantity $W$ involves derivatives of $\gamma_1$, it is 
even more sensitive to equation-of-state effects than $\gamma_1$
itself. That this is indeed the case is clearly reflected in 
Fig.~\ref{Fig11}, where the two models,
which are, after all, very close to each other, lead to values of $W$
which differ by an amount
far larger than the accuracy of the inversion (indicated
by error bars). Since the models of Fig.~\ref{Fig11}
exhibit exactly the same variation of the heavy-element composition
as the models of the present paper, it is clear that
Fig.~\ref{Fig11} has convincingly demonstrated
that the effects found and studied here are already
well within the reach of present-day observational accuracy.

\section{Discussion}

\subsection{Role of the occupation probability of the ground state}

As we have mentioned in section~1, the effect of the
treatment of excited states of atomic and ionic species can
show up under certain circumstances.
ND98 pointed out 
that the existence of the wiggle in the $\chi _\rho $ 
diagram in the MHD equation of state [see Fig. \ref{Fig1} (b)] is a 
genuine effect of neutral 
hydrogen even if it occurs in a region where 
most of hydrogen is already ionized. The wiggle is caused 
by the specific form of the 
density-dependent occupation probabilities of excited states in MHD.
However, ND98, did not examine 
the influence of the ground state.
Thus, to see if the excited 
states do, or do not, behave in
the same way as the ground states, we have carried out
a numerical experiment in which we have switched on and off 
the MHD-type occupation probabilities [see
Eq. (\ref{fo:w}) for definition] of {\it all ground states of all species}.
To switch off the MHD-type occupation probability of a ground state,
we have simply set $w_{1s}=1$, where $s$ stands for a given species
(atom or ion of an element), 
with all of the other $w_{is}$ remaining the same as usual.
Obviously, this is a purely academic exercise,
because such a choice of occupation probabilities 
is physically inconsistent. More specifically, by
robbing the ground-state occupation probability
of the possibility to become less than $1$, one disables
the capacity of the formalism to model pressure ionization.
However, here we merely intend to see what kind 
of role the occupation probability plays on the ground states and on the
excited states. A motivation for such a question is given by
the yet unexplained fact 
that the MHD equation 
of state is not as good as the OPAL equation of state for
temperatures between $\log T=5.4$ 
and $6$ inside the sun (see Fig. \ref{Fig10}).

For this test we have used a pure hydrogen-helium mixture.
For consistency with the work by ND98,
we have taken their run of temperature and density, which
corresponds roughly to the solar convection zone 
(here we refer to these conditions as ``solar track'',
to distinguish it from the more precise concept of a solar model
used in other figures). 
Our results are shown in Fig. \ref{Fig12}.
Besides the previously defined labels OPAL, CEFF, and MHD, here
the label
MHD$_{\rm{GS}}$ for the standard MHD internal 
partition function of hydrogen and helium truncated to the ground state 
term (see also ND98). 
MHD$_{\rm W1}$ and MHD$_{\rm GS,W1}$ are the same 
as MHD and MHD$_{\rm{GS}}$, respectively, except that in them
the occupation 
probabilities of all the ground states are set to be~$1$. 
The label MHD$_{\rm He:GS}$ refers to a truncation to the
ground state term of the helium internal partition function only. 
Similarly, MHD$_{\rm He:GS,W1}$ refers to
ground-states occupation probabilities in MHD$_{\rm He:GS}$ that are set 
to $1$.
We again stress that the absence of the possibility to model
pressure
ionization makes the MHD$_{\rm GS,W1}$ model quite unphysical.
Indeed, it is found to deviate significantly from all
other models at sufficiently high densities, where pressure ionization 
matter.

In Fig. \ref{Fig13}, 
we see that the ND98-wiggle between $\log T=4.5$ and $5.5$ 
shows up more distinctly
when
the occupation probability of the ground state is set to~$1$.
This clearly confirms the conclusion by ND98
that the wiggle must be a pure excited-states effect.
An occupation probability of the ground states different from~$1$ 
actually happens to reduce the wiggle. 
From Fig. \ref{Fig12} (a-c) we can see that 
MHD$_{\rm He:GS}$ model is very close to MHD model, and MHD$_{\rm 
He:GS,W1}$ is very close to MHD$_{\rm W1}$. The presence
of helium does not change very much, which is another confirmation of
the conclusion by ND98
that the wiggle is an excited-states effect of pure hydrogen.

Furthermore, it 
can be seen from Fig. \ref{Fig12} (a-c) that the effect of 
the ground-state occupation
probability shows 
up at temperatures $\log T > 4.5$, and it is most significant
for $5 < \log T < 6.6$. Although none of our results
appears to fit the OPAL equation of state closely (not surprising
with our academic exercise), nonetheless it follows that 
setting ground-state occupation probabilities to~$1$
does bring the results somewhat closer to OPAL.
It could well be that in the MHD occupation probability
formalism with $w_{is}$ of \citet{hum88}
[see Eq.(\ref{fo:w})],
the ground states might be too strongly perturbed.

\placefigure{Fig12}

\placefigure{Fig13}

\placefigure{Fig14}

Another interesting feature is shown in Fig. \ref{Fig12} (d) in
the temperature range 
$\log T=5.2$ to $\log T = 6.0$. By comparison
with Fig. \ref{Fig10}, we
see that 
the difference in 
$\gamma_1$ between MHD and MHD$_{\rm GS,W1}$ suspiciously
mimics
the difference between MHD and OPAL in that region.
Now, this is precisely the region where helioseismic inversions
[see, {\it e.g.} \citet{bdn99}] have given
evidence that the MHD equation of state is 
not as good as OPAL. In both cases, the intersection of both
MHD versions with OPAL happens at about the same
temperature ($\log T \approx 5.8$), 
and their slopes are about the same. This is an indication 
that, even if the extreme case of
$w_{1s}=1$ in the ground states can not be an overall
improvement of the MHD the equation of state,
the specific ground-state occupation numbers
$w_{is}$ of \citet{hum88}
[see Eq.(\ref{fo:w})] should be improved, most likely so that
they will be closer to~$1$.
Fig. \ref{Fig14} also tells us
that it is the choice of occupation probability of the ground states of
hydrogen that is responsible for the difference between MHD and OPAL 
from $\log T=5.5$ to $\log T = 6.0$.
This result shows another direction in which 
the MHD equation of state can be improved, namely by inclusion
of a more realistic pressure-ionization mechanism, likely
based on a hard sphere model [see, for instance, \citet{sch92}]. Such a
procedure could then
assure physical consistency even if $w_{1s}=1$.

Special attention is also in order for the cases
MHD$_{\rm GS,W1}$ and MHD$_{\rm He:GS,W1}$, in the
temperature range of $5.2 < \log T < 5.6$, shown 
in Fig. \ref{Fig12} (d). Both these cases
are close to OPAL. First, one can realize that 
it is the contribution of the excited states of helium that causes the 
behavior of the $\Delta 
\gamma_1$ line at the low-temperature end. 
Then, in this figure the wiggly feature from $4.0 < \log T < 5.2$
is once again the contribution of the excited states
of hydrogen. If we assume that OPAL is the better 
equation of state around the temperature of $\log T \approx 5.8$ 
then the occupation probability expression used by \citet{hum88} might
indeed
have altered the 
partition function of the ground states too strongly. 

\subsection{Effect of the $\tau$ correction}

Another numerical
experiment was dedicated to
the validity of the $\tau$ correction to the Debye-H\"uckel term,
such as employed in the MHD equation of state, but also in CEFF. As 
mentioned in section~1, the $\tau$ correction is 
conventionally added \citep{gra69}
to remedy the effects of the divergence of the Debye potential
at $r=0$. The net effect of the $\tau$
correction is to prevent the negative Debye-H\"uckel pressure 
correction from exceeding
the ideal-gas pressure which would, at very high densities, 
cause a negative total 
pressure. In Fig. \ref{Fig15}, we have plotted the value 
of $\tau$ for our solar model. 
Because the Debye-H\"{u}ckel correction reaches its maximum in the middle of 
the solar convection zone,
$\tau$ is there also bigger than elsewhere.
In parallel, in that region
$\chi_\rho$ is enhanced and $\chi_T$ reduced with respect to the 
other equations of state that have no 
$\tau$ correction (Fig. \ref{Fig16}). 

A comparison of 
Fig. \ref{Fig17} (a-c) with Fig. \ref{Fig15} makes one realize
that the
$\tau$ contribution shows up clearly 
in pressure, $\chi_\rho$ and $\chi_T$,
revealing how these thermodynamic quantities
differ from those of equations of state without a $\tau$ correction. 
The $\tau$ correction leads to a significant change of the 
thermodynamic quantities
in the solar convection zone. If one assumes that OPAL (which does
not contain a $\tau$ correction) is in all respects quite close
to the true equation of state in this temperature range,
one can conclude that a $\tau$ correction would lead to
inconsistent quantities $\chi_\rho$ and $\chi_T$ .
However, in $\gamma_1$ [Fig. 
\ref{Fig17} (d)], the behavior of the $\tau$ correction 
is more complicated and concealed.
That is certainly one of the reason why helioseismic studies so far have
not had problems with the $\tau$ correction. 
For instance, very successful solar models have been constructed
with CEFF and MHD.
As a word of caution we mention,
however,
that the recipe of improving the MHD equation of state by
simply removing its $\tau$ correction would not work in all stellar
applications. For the sun it does work, because
nowhere inside does the Debye-H\"uckel
term come even close to cause negative pressure.
In contrast, an application of
the MHD equation of state to the physical conditions of low-mass stars
would be confronted with 
this pathology, and adding the $\tau$ correction is a must, already
for formal reasons, independent of physical merit.
Incidentally, the MHD equation of state with a 
$\tau$ correction has turned out to be a useful working
tool for low-mass stars~\citep{cha99},
but this might have been due to fortuitous circumstances.
For a more realistic physical description,
high-order contributions in the Coulomb interaction beyond the Debye-H\"uckel 
theory will be needed.

\placefigure{Fig15}

\placefigure{Fig16}

\placefigure{Fig17}

\section{Conclusions}

\begin{figure}
\figcaption{Difference in thermodynamic quantities  
between the H-He-C mixture and the H-He mixture for various
equations of state.  Differences are 
in the sense [X(Model)$_{\rm H-He-C}$ - X(Model)$_{\rm H-He}$]. 
Line styles the same as in Fig. (\ref{Fig1}).
\label{Fig6}}
\plottwo{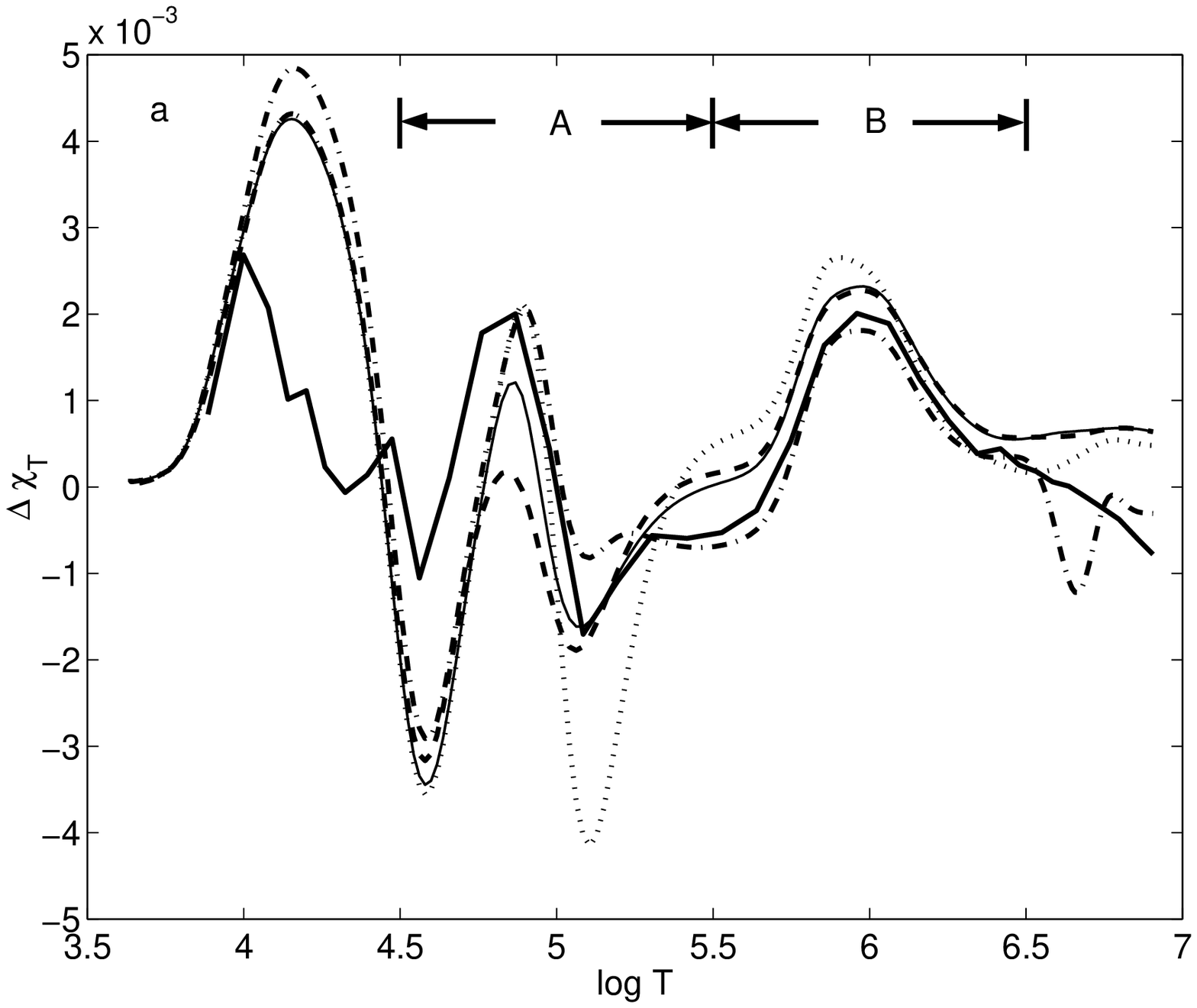}{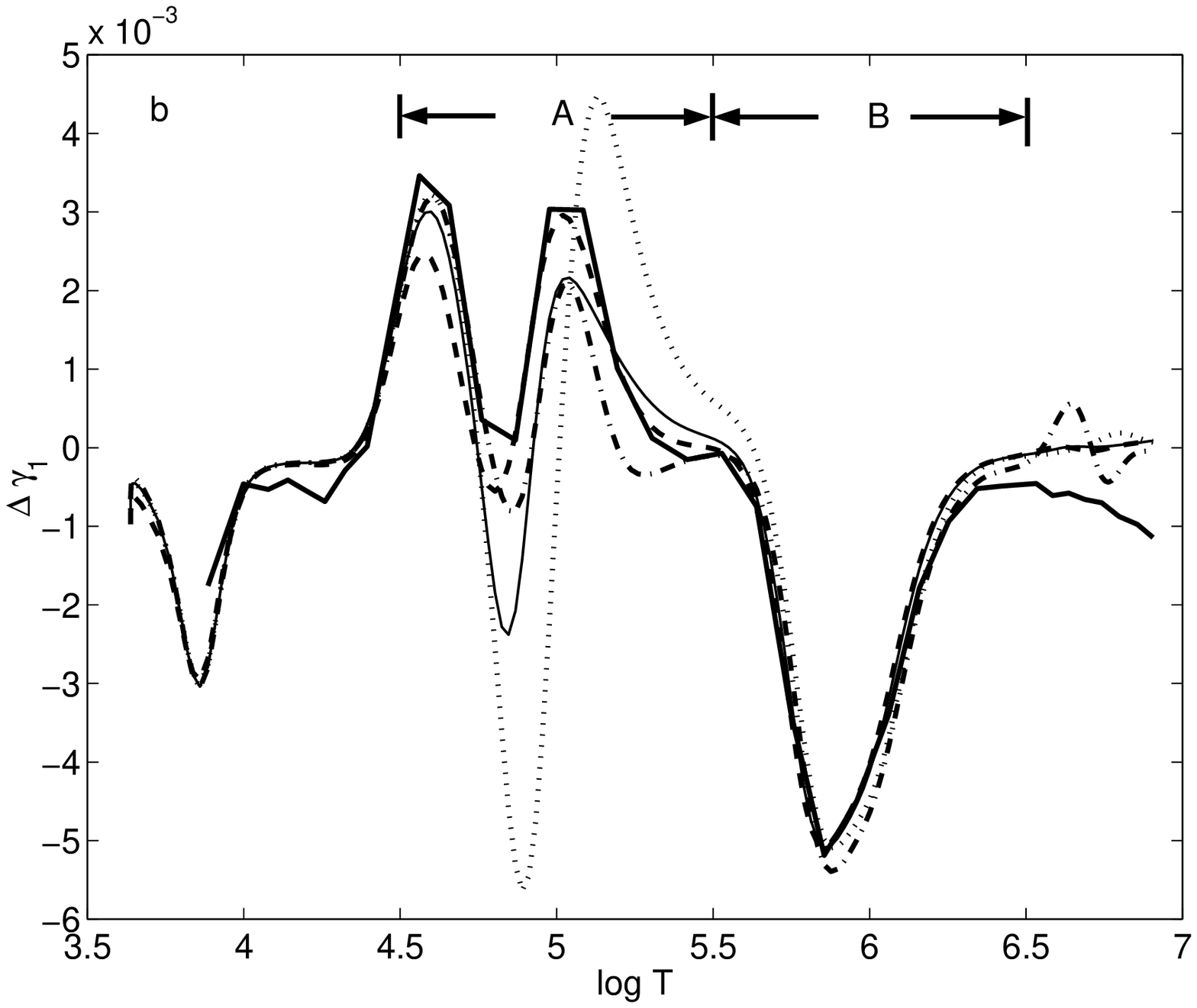}
\end{figure}

\begin{figure}
\figcaption{Difference in thermodynamic quantities  
between the 6-element mixture and the H-He-C mixture for various
equations of state. Differences are 
in the sense [X(Model)$_{\rm 6-mixture}$ - X(Model)$_{\rm H-He-C}$]. 
Line styles the same as in Fig. 
(\ref{Fig1}).
\label{Fig7}}
\plotone{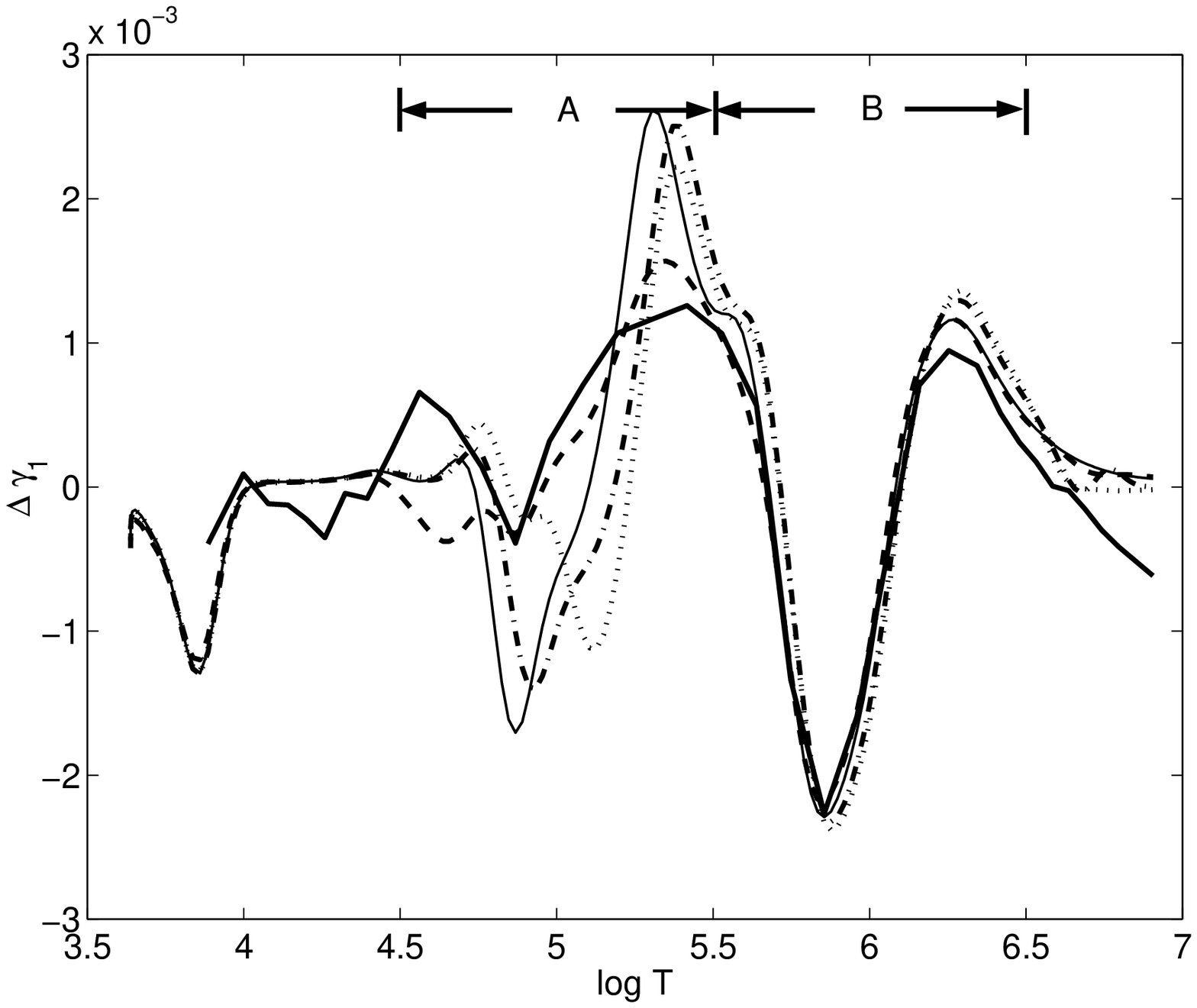}
\end{figure}

\begin{figure}
\figcaption{Difference in thermodynamic quantities  
between the H-He-N mixture and the H-He mixture for some models of 
equation of state. Differences are in the sense
[X(Model)$_{\rm H-He-N}$ - X(Model)$_{\rm H-He}$]. 
Line styles the same as in Fig. (\ref{Fig1}).
\label{Fig8}}
\plotone{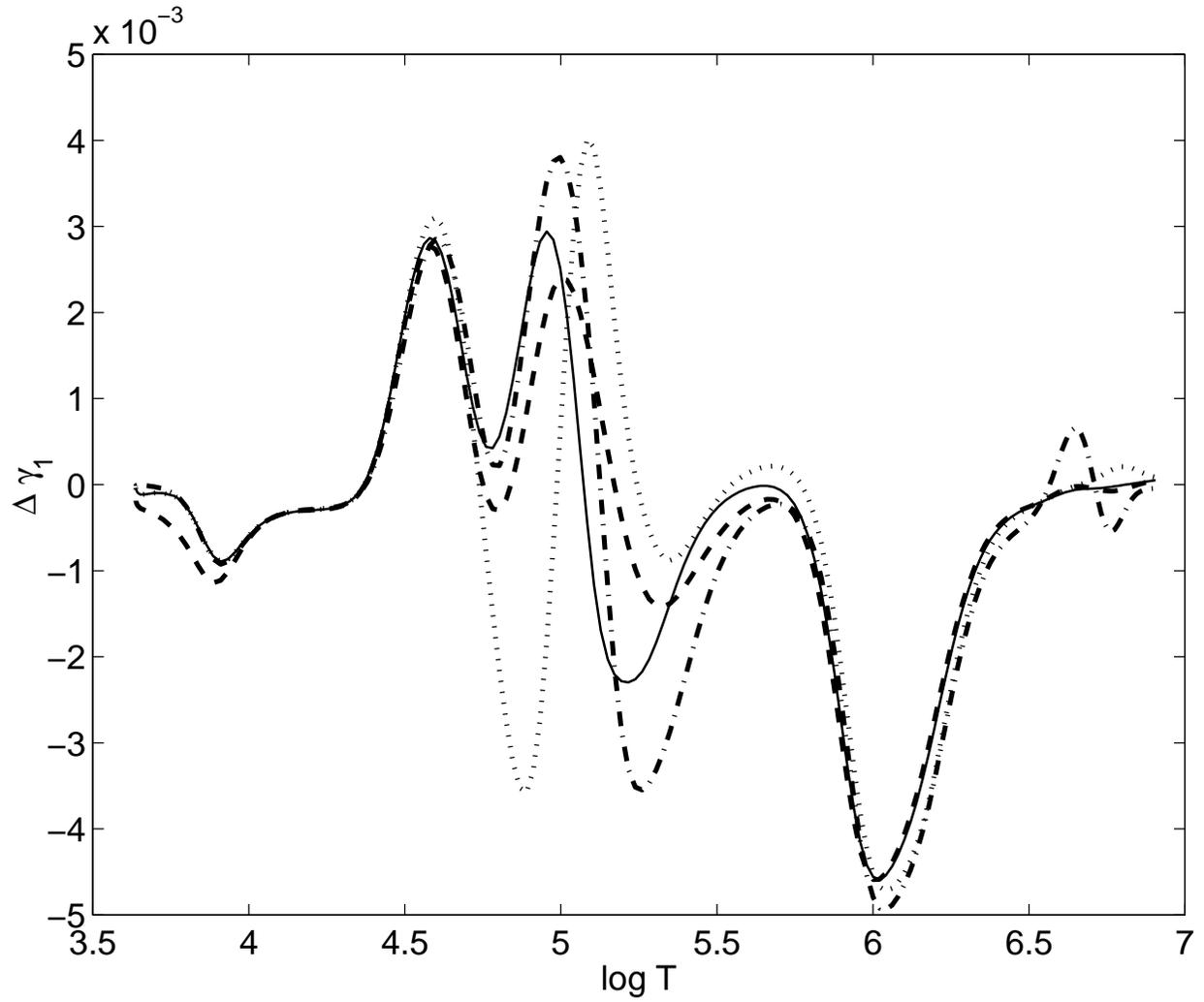}
\end{figure}

\begin{figure}
\figcaption{Relative difference of thermodynamic quantities  
between a sequence of reduced mixtures and the full 
15-element mixture for the MHD equation of 
state. Differences are in the sense 
(X - X$_{\rm 15-element}$)/X$_{\rm 15-element}$. 
In panels (a-c): Thin Solid line: 4-element; Dashed-dotted
Line: 5-element; Dashed Line: 6-element; Thick Solid Line: 7-element; Dotted
Line: 11-element. In panel (d): Thin Solid line: 3-element; Dashed-dotted
Line: 4-element; Dashed Line: 5-element; Thick Solid Line: 6-element; Dotted
Line: 7-element.
\label{Fig9}}
\plotone{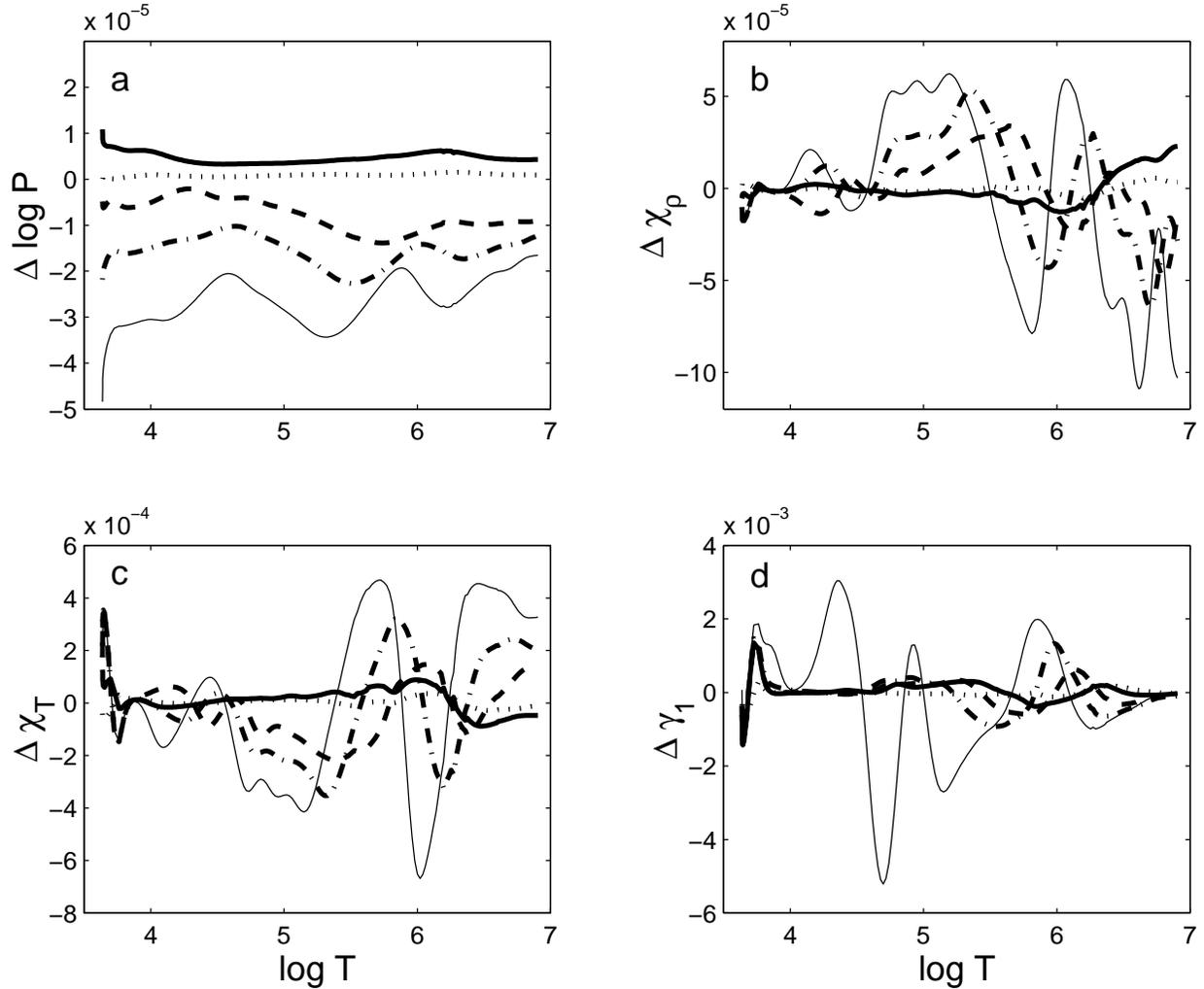}
\end{figure}

\begin{figure}
\figcaption{
Intrinsic difference between $\gamma_1$ obtained from an inversion of
helioseismological data~\citep{bdn99}, 
and $\gamma_1$ of the solar models M1--M8 listed
in Table~\ref{Tb2}, 
in the sense ``sun -- model''.
The intrinsic difference is due to the change in the equation of
state alone (see text).
Filled points label results
from MHD models, empty ones OPAL.
Lines have been drawn through results of models M1 and M5 to guide the
eye. For the sake of clarity, 
error bars have been drawn only on two sets of results.
\label{Fig10}}
\plotone{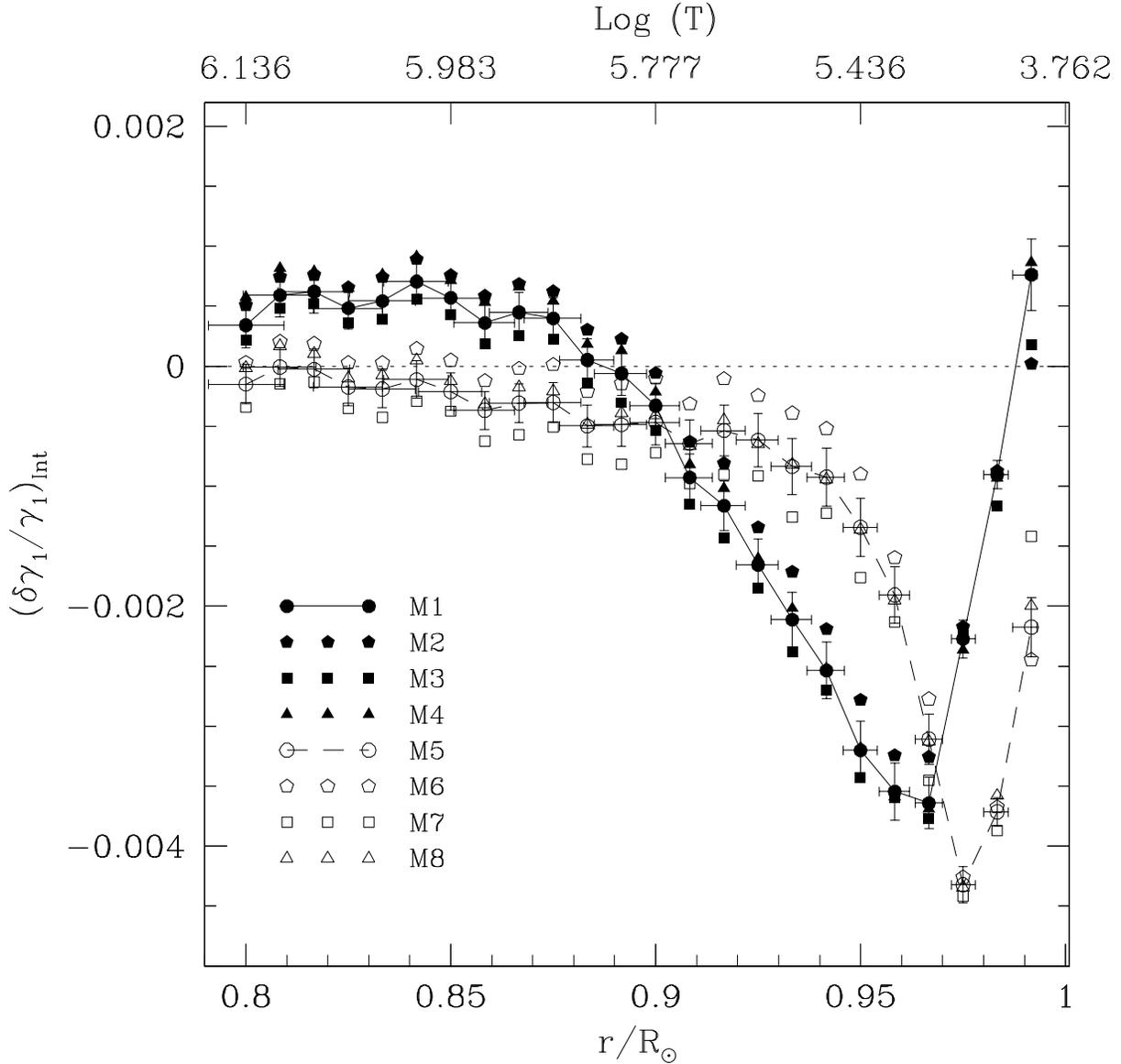}
\end{figure}

\begin{figure}
\figcaption{The quantity W=$({\rm d}c^2/{\rm d}r)/g(r)$ obtained
from an inversion of solar p-mode frequencies and three
solar models with different heavy-element abundances~\citep{bat00}.
The observational curve is marked with vertical bars (estimated 
inversion error). The pair of solid and dashed lines that
follows the observational $W$ most closely is from
a model with
the usual solar composition (dashed line: $W$ computed
directly from the solar model; solid line: result of
the same inversion procedure as for the observational
curve, but with artificial mode frequencies from the solar model).
The two other pairs of solid and dashed lines 
(``H-He-C'' and ``H-He-O'') are from 
models with a heavy-element composition of pure
carbon and oxygen, respectively, of an amount equal to
the total solar heavy-element abundance. Solid and dashed lines
have the same meaning as before.
In contrast to Fig. \ref{Fig10}, the inversions here are
absolute (that is, not relative to a reference model).
All models in this figure include
electrostatic screening (Eq.~\ref{eq:scr}).
\label{Fig11}}
\plotone{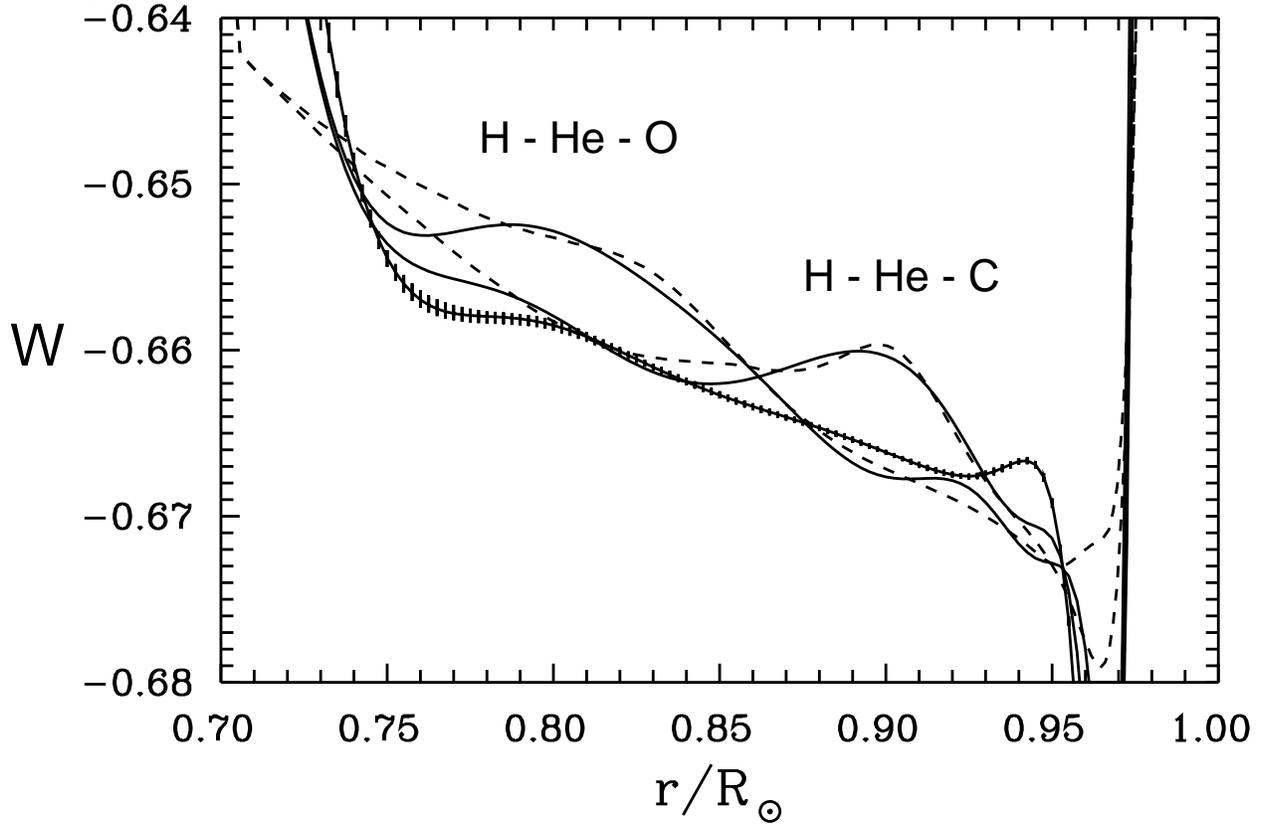}
\end{figure}

\begin{figure}
\figcaption{Relative difference between thermodynamic
quantities for the H-He mixture in a test with setting
the occupation probability of the ground states $w_{1s}=1$ (see text
for a description). 
Differences are in the sense 
(X - X$_{\rm MHD_{\rm GS}}$) / X$_{\rm MHD_{\rm GS}}$. 
Thick Dashed Line: MHD; Thick Dashed-dotted Line: OPAL; Dotted Line: CEFF; 
Thin Dashed-dotted Line: MHD$_{\rm W1}$; Thin Dashed Line: MHD$_{\rm GS,W1}$;
Thin Solid Line: MHD$_{\rm He:GS}$; Thick Solid Line: MHD$_{\rm He:GS,W1}$.
\label{Fig12}}
\plotone{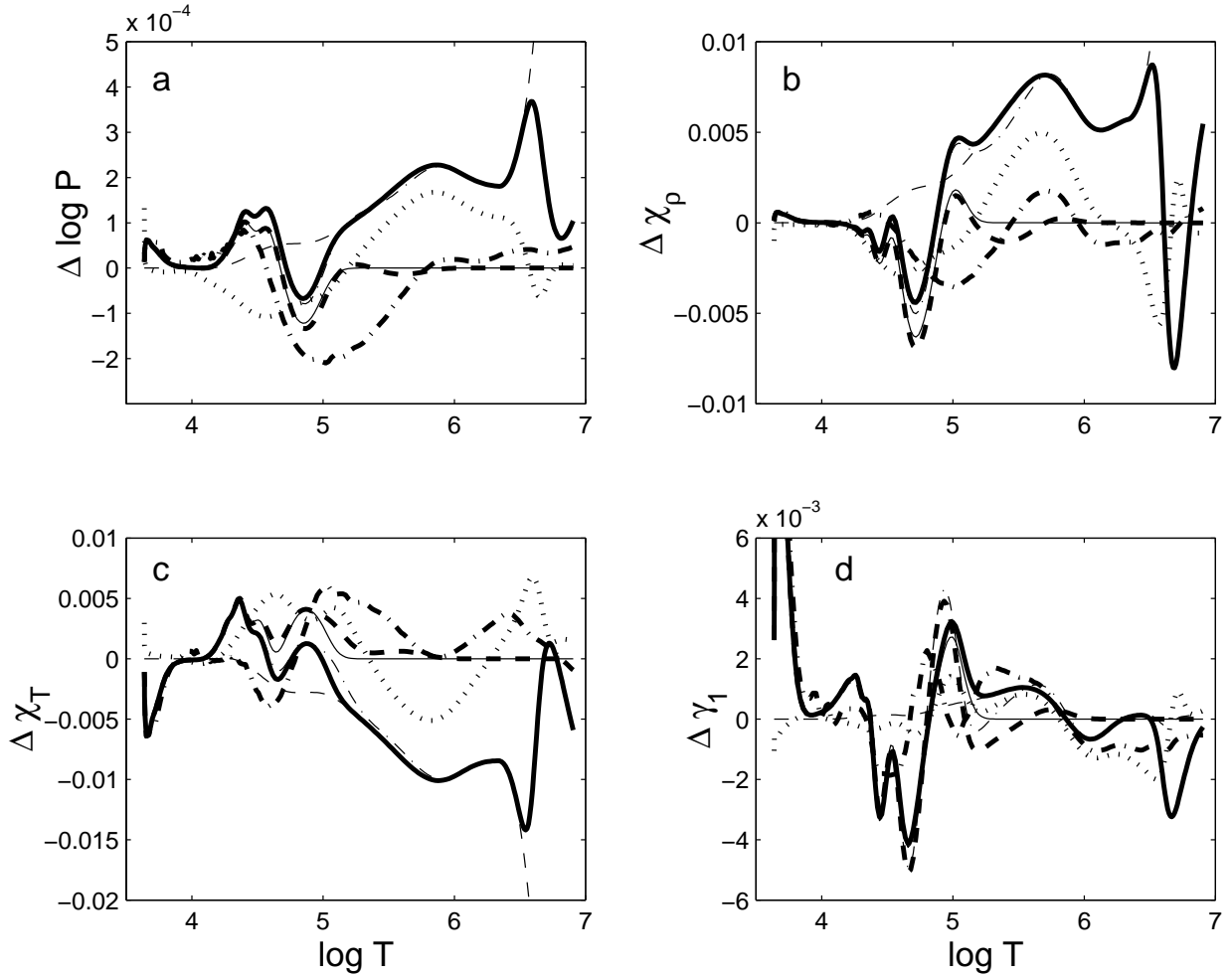}
\end{figure}

\begin{figure}
\figcaption{Absolute value of $\chi_\rho$ for the H-He 
mixture in a test with setting
the occupation probability of the ground states $w_{1s}=1$ (see text
for a description).
Thick Dashed Line: MHD; Thick Solid Line: OPAL; Dashed-dotted Line: 
CEFF; Dotted Line: MHD$_{\rm W1}$; Thin Solid Line: MHD$_{\rm GS,W1}$;
Thin Dashed Line: MHD$_{\rm GS}$. 
\label{Fig13}}
\plotone{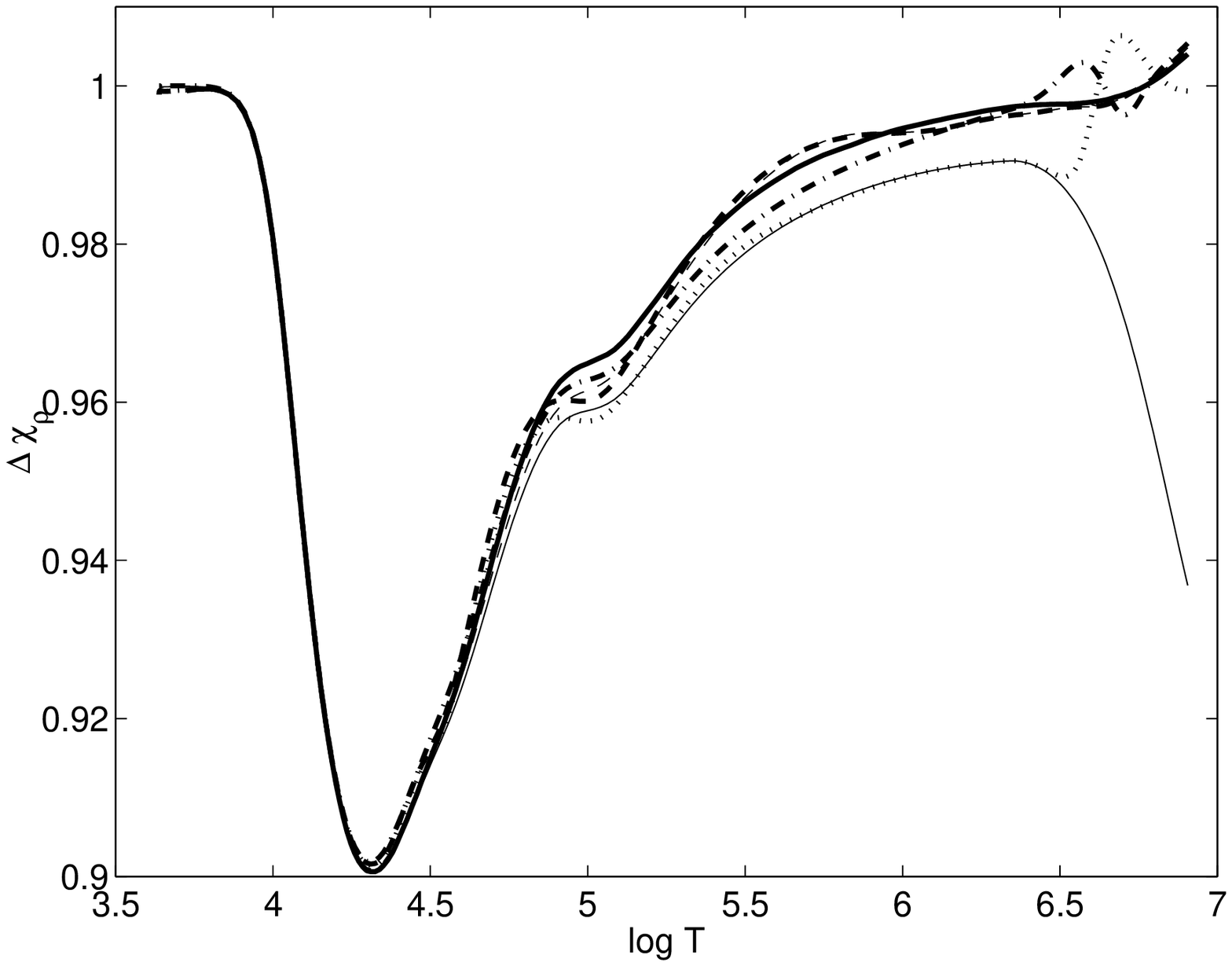}
\end{figure}

\begin{figure}
\figcaption{Relative difference between thermodynamic
quantities for the pure hydrogen plasma in a test with setting
the occupation probability of the ground states $w_{1s}=1$ (see text
for a description).
Differences are in the sense
(X - X$_{\rm MHD_{\rm GS}}$) / X$_{\rm MHD_{\rm GS}}$. 
Dashed Line: MHD; Dashed-dotted Line: CEFF; Dotted Line: MHD$_{\rm 
W1}$; Solid Line: MHD$_{\rm GS,W1}$.
\label{Fig14}}
\plotone{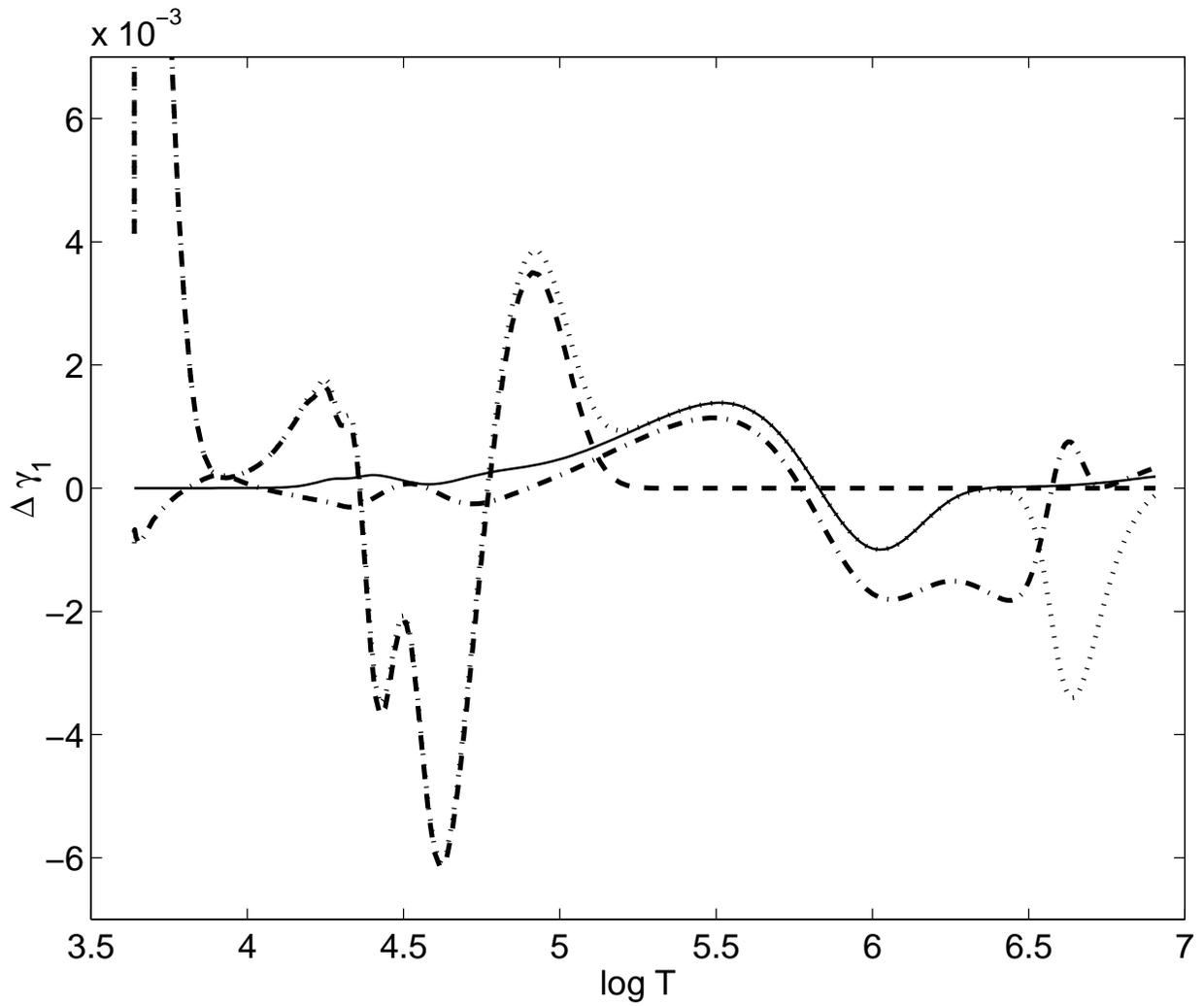}
\end{figure}

\begin{figure}
\figcaption{The $\tau$ correction in the solar convection
zone.
\label{Fig15}}
\plotone{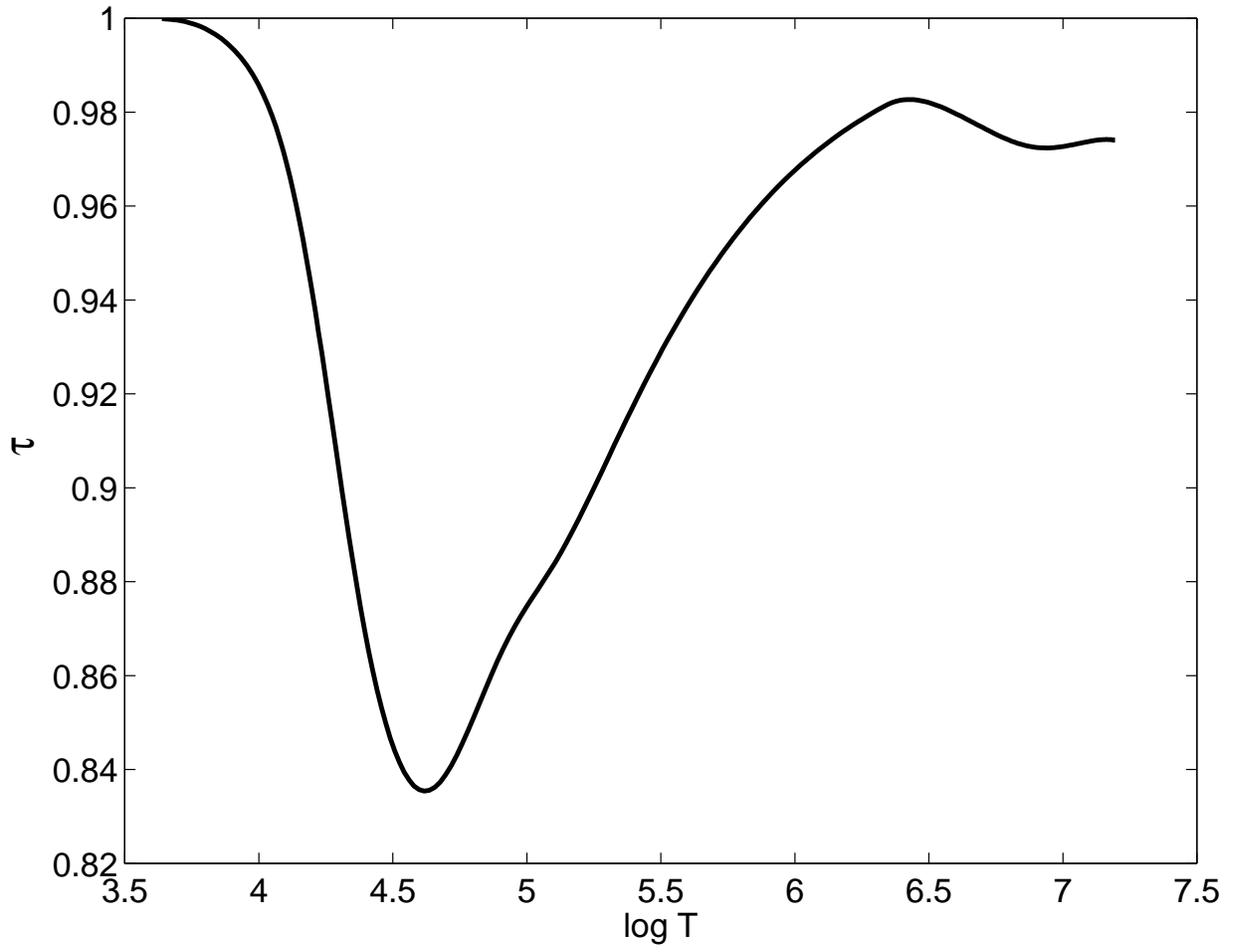}
\end{figure}

\begin{figure}
\figcaption{Values of $\chi_\rho$ and $\chi_T$ for the H-He 
mixture with and without $\tau$ correction inside the sun. 
Solid Line: MHD; Dotted 
Line: MHD$_{\rm GS}$; Dashed Line: OPAL; Dashed-dotted Line: MHD with $\tau$ 
correction.
\label{Fig16}}
\plottwo{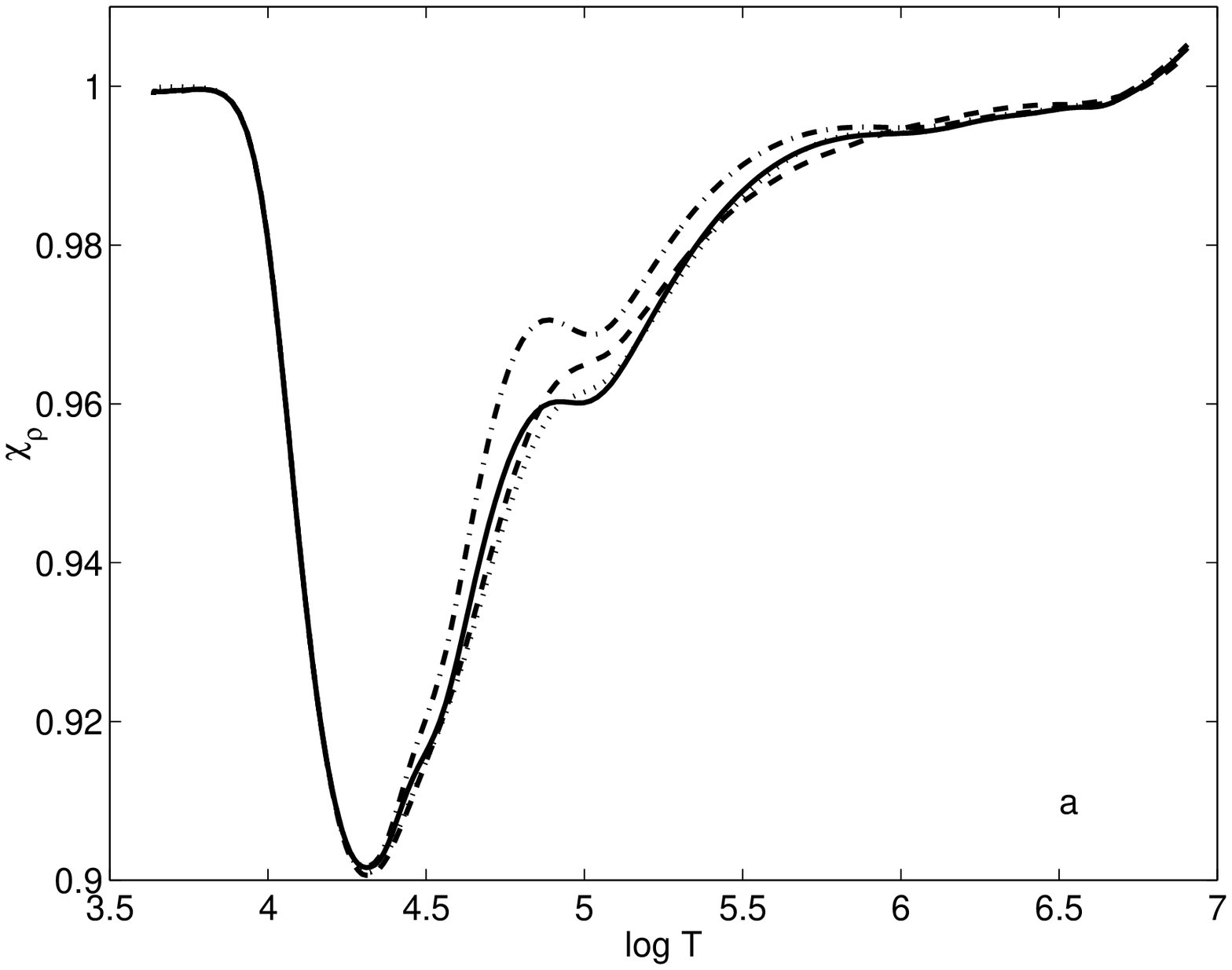}{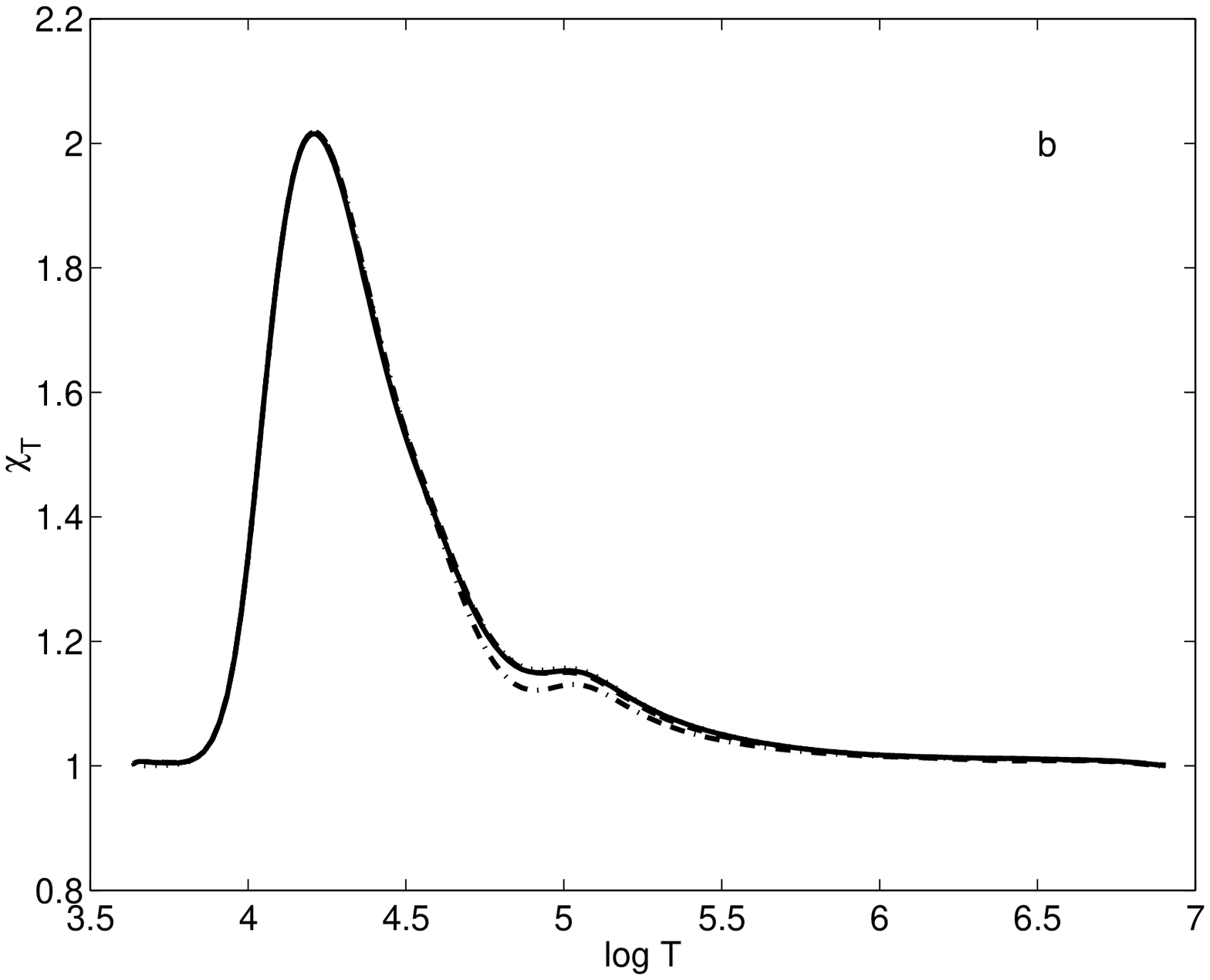}
\end{figure}

\begin{figure}
\figcaption{Relative difference of thermodynamic
quantities for the H-He mixture with 
and without $\tau$ correction inside the sun. 
Differences are in the sense
(X - X$_{\rm MHD_{\rm GS}}$) / X$_{\rm MHD_{\rm GS}}$. 
Solid Line: MHD; Dashed Line: OPAL; Dashed-dotted Line: MHD with 
$\tau$ correction.
\label{Fig17}}
\plotone{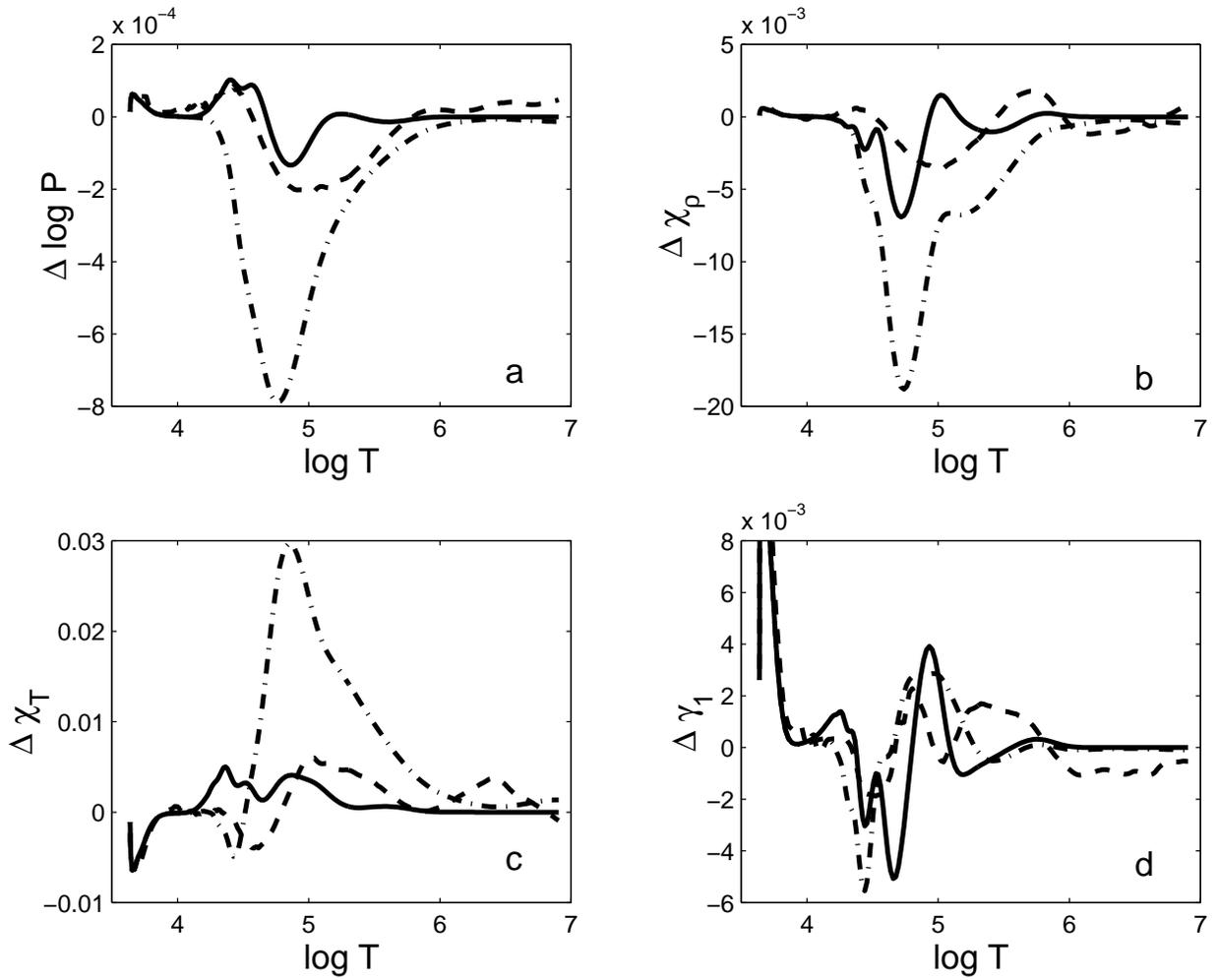}
\end{figure}

The first part of the paper has been dedicated to
the influence of heavy elements in thermodynamic quantities. 
To isolate the contribution of a selected heavy elements separately,
we have compared the results from a H-He-C mixture with those of 
a H-He mixture. It has emerged that for
temperatures between $4.5 < \log T <5.5$ (region ``A'' in Fig. \ref{Fig6}
and Fig. \ref{Fig7}), the thermodynamic quantities
are very sensitive to the detailed physical 
treatment of heavy elements, in particular regarding
the excited states of all atoms and ions of the
heavy elements. These findings carry an important diagnostic
potential to use the sun as a laboratory to constrain
physical theories of atoms and ions immersed in hot dense plasmas.

However, we have also found that
not all physical effects of heavy-elements can be detected
with helioseismic studies. The reason is that
in a realistic mixture with many heavy elements,
at certain places in the solar convection zone, 
the profile of the
relevant adiabatic gradient
is significantly smoother than for less realistic 
artificial mixtures which contain a smaller number
of heavy elements. As a consequence, in some locations,
the adiabatic gradient can become quite
independent of details of the physical
treatment of the species. 
Such a property is welcome news for solar modelers, because it reduces
the uncertainty due to the equation of state. A prime
beneficiary of this enhanced precision will be the helioseismic
helium and heavy-element
abundance determination.
To this purpose, we have identified in the adiabatic gradient 
a useful device in the form of 
a prominent, largely model-independent
feature of heavy elements. This feature is found
at temperatures around $5.5 < \log T < 6.5$ (region 
``B'' in Fig. \ref{Fig6} and Fig. \ref{Fig7}), corresponding
roughly
to
the base of solar convection zone, where 
each heavy element exhibits its own
ionization profile. We verified that these profiles are
indeed quite independent
of the details in the physics. They can serve as
tracers for heavy elements, 
and they harbor the
potential for a
helioseismic determination of
the relative abundance of heavy elements in the
solar convection zone.

The second part of the paper deals with related issues.
It is no surprise that 
the major contribution of the
heavy elements to pressure
is given by the total number particles involved.
These are nuclei and the electrons
released by ionization, which is mainly determined
by temperature and 
the relevant ionization energies.
Somewhat less expected is the result that with a larger number of heavy
elements, the profile of 
thermodynamic quantities becomes smoother than in the case
of a low number (one or two) representative
heavy elements. In a quantitative study, we found that
6-element mixtures, which are still widely used in 
solar modeling, may contain errors
in $\gamma_1$ of up to $3 \times 10^{-4}$ due to the insufficient
number of heavy elements.
Such a discrepancy does matter in
present helioseismic studies (section~3.4). 
We conclude that in order to avoid this error, the element mixture must 
contain at least
10 of the most abundant elements.

In a third part, rather as a by-product of the present study, 
we discovered one important reason
why in helioseismic studies, the MHD equation of state does 
not produce as good results as the OPAL equation of state. 
We developed an unphysical diagnostic formalism,
where the occupation 
probability of the ground states of all species was
left untouched at~$1$. With this simple tool,
we have on the one hand confirmed the  
conjectures of ND98 about the importance of
the ground-state contribution compared to that of the excited states.
On the other hand, we have also shown that the difference 
between the MHD and OPAL equations of state around 
$\log T \approx 5.8$ appears to be due to MHD's specific choice
of the occupation probability of the ground state of hydrogen. 
We have realized that the ground-state contribution 
clearly moves away the MHD model both from
helioseismologically determined values and OPAL.
This result suggests that
the specific occupation probability adopted by \citet{hum88} might perturb
the ground states (and perhaps also the low-lying excited states) 
too strongly.

In a final part, and as an another by-product, 
we have obtained quantitative
results about the effect of the $\tau$ correction term, which 
is sometimes added to
Debye-H\"uckel theory. We confirm earlier conjectures~\citep{bat95} that
the $\tau$ correction causes a significant spurious effect in 
the solar equation of state
and is therefore inadmissible. The $\tau$ correction should
therefore be taken out of the 
MHD equation of state (and CEFF for that matter). Such a remedy
will be acceptable for solar applications, because of
the overall smallness of the Debye-H\"uckel correction in
the sun. However, some
form of a 
$\tau$ correction is still be required 
for low-mass stellar modeling 
with the MHD and CEFF equations of state, because it has to
prevent the total
pressure from becoming negative at high densities and relatively low
temperatures. 
We note in passing that 
the OPAL equation of state does not need a $\tau$
correction because it contains genuine higher-order Coulomb correction
terms. The major discrepancy between MHD and OPAL -- other than the
aforementioned ground-state effect -- is not an absence of higher-order
Coulomb terms in MHD, but the presence of incorrect ones in the
form of the $\tau$ correction. The MHD equation of state 
should be upgraded
to include higher-order Coulomb contributions.

\acknowledgments

We thank Forrest Rogers for stimulating discussions and most of the ACTEX 
equation of state data used in this study, J\o rgen Christensen--Dalsgaard for 
the solar model used in this study, Alan Irwin and Joyce Guzik for the
SIREFF program from which some of the comparison data are generated. This
work was supported by the grants AST-9618549 and AST-9987391 of the 
National Science Foundation and the SOHO Guest Investigator grants 
NAG5-7352 and NAG5-7902 of NASA. 
SOHO is a project of international cooperation 
between ESA and NASA.

\clearpage

\begin{table} 
\caption{Chemical composition of the heavy elements in the 
6-element mixture} \label{Tb1}
\begin{center}
\begin{tabular}{lcc}
\tableline \tableline
Element & Relative Mass Fraction & Relative Number Fraction\\
\tableline
Carbon & 0.1906614 & 0.2471362\\
Nitrogen & 0.0558489 & 0.0620778\\
Oxygen & 0.5429784 & 0.5283680\\
Neon & 0.2105114 & 0.1624178\\
\tableline
\end{tabular}
\end{center}
\tablecomments{This composition is employed in current OPAL 
tables~\citep{rog96}.}
\end{table}

\begin{table}

\caption{Properties of the solar models used in
Fig.~\ref{Fig10} (see text). 
\label{Tb2}}

\begin{center}
\begin{tabular}{llllll}
\hline

Model & \hbox{Equation} & \hbox{Radius} & \hbox{Convective}& 
  $Y_{\rm s}$& $r_{\rm cz}/R_\odot$ \\
  & \hbox{of State} & \hbox{Mm} & \hbox{Flux} & & \\

\hline

M1 & MHD & 695.78 & CM & 0.2472 & 0.7145 \\
M2 & MHD & 695.99 & CM & 0.2472 & 0.7146 \\
M3 & MHD & 695.51 & CM & 0.2472 & 0.7145 \\
M4 & MHD & 695.78 & MLT & 0.2472 & 0.7146 \\
M5 & OPAL& 695.78 & CM & 0.2465 & 0.7134 \\
M6 & OPAL& 695.99 & CM & 0.2465 & 0.7135 \\
M7 & OPAL& 695.51 & CM & 0.2466 & 0.7133 \\
M8 & OPAL& 695.78 & MLT & 0.2465 & 0.7135 \\

\hline
\end{tabular}
\end{center}
\end{table}

\clearpage

\end{document}